\documentclass[10pt,journal,onecolumn]{IEEEtran}

\usepackage{filecontents}
\usepackage{lipsum}
\usepackage{microtype}
\usepackage{lipsum}
\usepackage{filecontents}
\usepackage{lipsum}
\usepackage{umoline}

\usepackage{subcaption}
\usepackage{caption}
\usepackage{multirow}

\ifCLASSINFOpdf

\else

\fi

\usepackage[most]{tcolorbox}

\newtcolorbox{myframe}[1][]{
	enhanced,
	arc=0pt,
	outer arc=0pt,
	colback=white,
	boxrule=0.8pt,
	#1
}
\usepackage{lipsum,graphicx,subcaption}

\usepackage{bbding}
\usepackage{pifont}
\usepackage{wasysym}
\usepackage{amssymb}

\usepackage[english]{babel}
\usepackage{scrhack}
\usepackage{graphicx}
\usepackage{wrapfig}
\usepackage[utf8]{inputenc}
\usepackage{lipsum}
\usepackage{cuted}
\usepackage{amsfonts}
\newtheorem{Theorem}{Theorem}

\newtheorem{Definition}[Theorem]{Definition}

\newtheorem{Remark}[Theorem]{Remark}

\lstdefinestyle{Normaltext}{language=,numbers=none,basicstyle=\normalfont}

\lstnewenvironment{textbox}[2]
{%
	\centering
	\minipage{1\textwidth}
	\lstset{style=Normaltext,label=#1,caption=#2}
}
{%
	\endminipage
	\endcenter
}
\usepackage{tempora}

\usepackage[
pdftitle={Math Assignment},
pdfauthor={Joe Doe, Some University},
colorlinks=true,linkcolor=blue,urlcolor=blue,citecolor=blue,bookmarks=true,
bookmarksopenlevel=2
]{hyperref}

\usepackage{tikz}

\newcommand*{\myalign}[2]{\multicolumn{1}{#1}{#2}}

\usepackage{lipsum}
\usepackage{mwe}

\usepackage{graphicx}
\usepackage{mwe}
\usepackage{hyperref}
\hypersetup{pdftoolbar=true,pdfpagemode=UseNone,pdfstartview=FitH,
	colorlinks=true,extension=}

\newcounter{multifig}

\newcommand{\multifig}{\setcounter{multifig}{\thefigure}}

\newcommand{\figtarget}{\refstepcounter{figure}}

\newcommand{\figcaption}[1]
{\stepcounter{multifig}
	\addcontentsline{lof}{figure}{\string\numberline {\arabic{multifig}}{\ignorespaces #1}}
	Figure \arabic{multifig}: #1}

\ifCLASSOPTIONcompsoc
  \usepackage[nocompress]{cite}
\else
  \usepackage{cite}
\fi
\ifCLASSINFOpdf
\else
\fi
\begin{document}
%
\title{Federated Online/Offline Remote	Data Inspection for Distributed Edge Computing}
%
%
%
%

\author{Mohammad~Ali
	and  Ximeng~Liu
	
	\thanks{M. Ali is  with the Department of Mathematics and Computer Science, Amirkabir University of Technology, Tehran, Iran. E-mail: mali71@aut.ac.ir}
	\thanks{X. Liu is with the Key Laboratory of Information Security of Network Systems, College of Computer and Data Science, Fuzhou University,
		Fuzhou 350108, China. Email: snbnix@gmail.com.  
}}

%
%

\markboth{ }%
{Ali \MakeLowercase{\textit{et al.}}: Federated Online/Offline Remote Data 	Inspection for Distributed Edge Computing}
%



\IEEEtitleabstractindextext{%
\begin{abstract}
In  edge computing  environments, app vendors can cache their data to be shared with their users in many  geographically distributed edge servers. However, the cached data is particularly vulnerable to several  intentional attacks or unintentional events.  Given the limited  resources of edge servers and prohibitive  storage costs incurred by app vendors, designing an  efficient  approach to inspect and maintain  the  data over tremendous edge servers is a critical  issue. 
To tackle the  problem, we  design a novel data inspection approach, named  ${\text{O}^2\text{DI}}$,  that provides the following  services: i) using ${\text{O}^2\text{DI}}$,   app vendors can inspect the  data cached in edge servers without having the original data, which reduces the incurred I/O and storage overhead significantly; ii) computational operations  conducted  by both edge servers and app vendors are highly efficient because of  a novel online/offline technique; iii) many data files cached in different edge servers can be verified  quickly and at once by using a novel batch verification method;  iv) corrupted data in edge servers can be  localized and   repaired efficiently. We analyze  the security and performance of ${\text{O}^2\text{DI}}$. We see that it is secure in the random oracle model,   much faster, and  more cost-effective compared  to state-of-the-art approaches.
\end{abstract}

}

\maketitle

\IEEEdisplaynontitleabstractindextext

%
\IEEEpeerreviewmaketitle

\section{Introduction}
With the rapid   growth of mobile and Internet of Things (IoT) technologies, the use of online applications such as networked gaming,  real-time video analysis, and virtual/augmented reality   has become ubiquitous  in this era of digital transformation \cite{A1}. At first glance, one may be tempted to think that  the  cloud computing paradigm is a  suitable tool to directly provide users of such  applications with the required services  in a scalable manner. However, as  the latency between the remote cloud server and end-users might be
high and  unpredictable, and  online applications are   highly sensitive to network latency, the conventional cloud computing paradigm   has been rendered obsolete for this purpose \cite{A2}. Therefore,  app vendors providing  online application services are in urgent need of  alternative distributed computing models with low latency.  To fulfill this  requirement, the distributed edge computing (EC)  environment has been put forward as a  complement to the cloud computing paradigm and also  a facilitator  of the 5G network \cite{A3}.

\begin{figure}[h]
	\centering
	\includegraphics[width=2.8in]{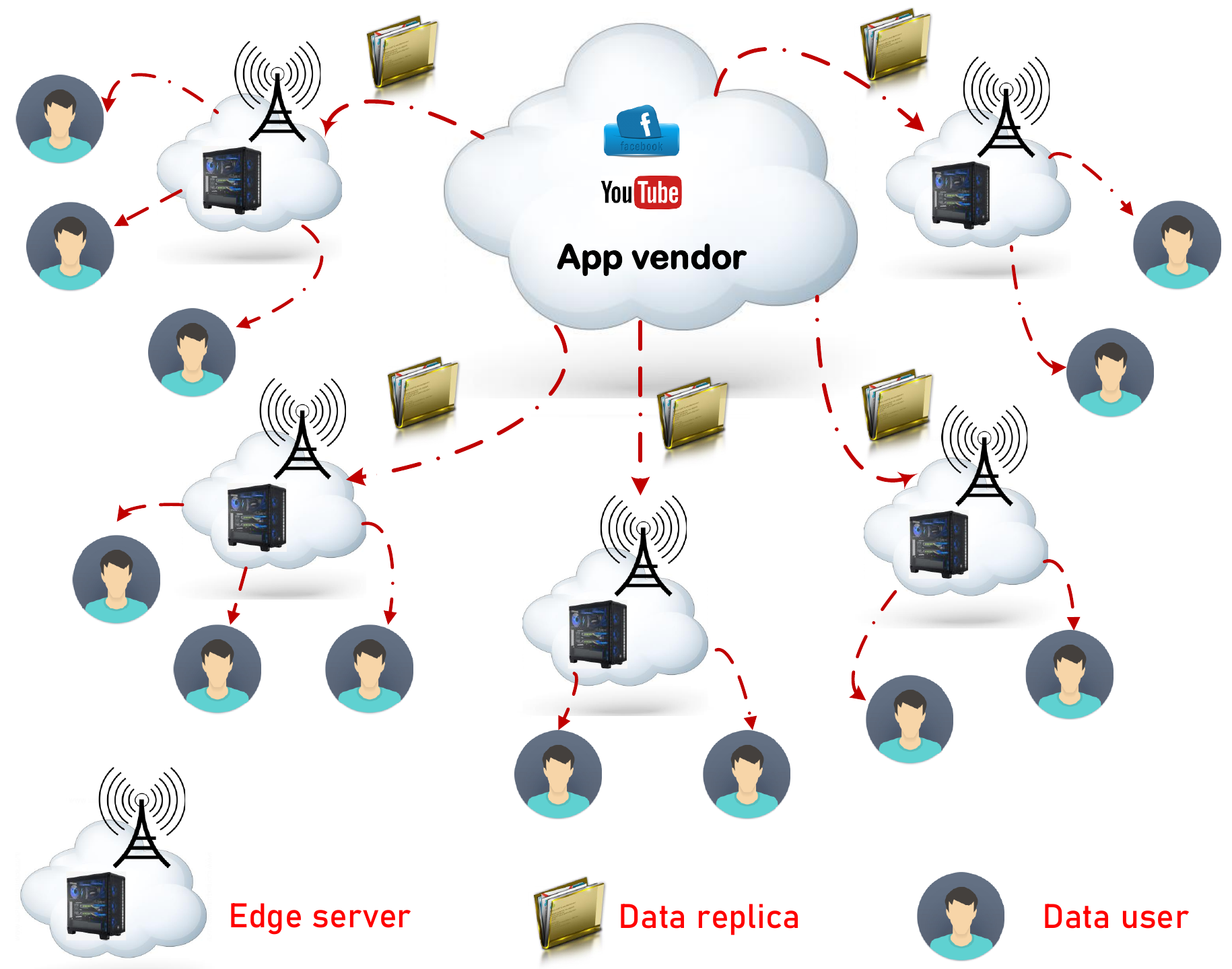} 
	\caption{A typical EC-based caching system. }
	\label{General}
\end{figure}

In  EC environments, the edge infrastructure provider  deploys several edge servers (ESs) in close proximity to end-users. ESs are usually  equipped with small-scale cloud-like computing and storage capabilities at access points and base stations \cite{A4}. 
To alleviate  the latency problem,  app vendors can hire some ESs  in  different  geographical areas to  host online applications  and cache  required data  to serve their users in those  areas with  high efficiency and low latency \cite{A5,A6}.  More precisely, in a typical EC-based  environment, when an app vendor   wants to make a data file  $M$ available to its users, it can generate  replicas  $M^{(1)},\ldots,M^{(N)}$  of $M$ and cache each of them to an ES (see Figure \ref{General}).   In this way, the app vendor can fulfill the  stringent latency requirement.

\textit{Challenge 1: Data integrity.}
It has been frequently   noted in the   literature \cite{A8,A9,alii2,alii3} that the data outsourced to a cloud  server might be subject to various  attacks aiming to corrupt or tamper with the data. Therefore,  in almost all  security models,  the cloud server is not   trustworthy and reliable \cite{A7,A10,alii4,alii5}. A similar situation holds for EC servers. Indeed, most of  attacks  on cloud computing-based systems  can also be applied to  edge computing-based environments. Moreover,  unlike cloud servers that usually have   abundant computing resources, edge servers possess limited  resources to protect their cached data against cyberattacks and other malicious behaviors \cite{LI,A11}. Therefore, data integrity in EC environments is at least as important as in cloud computing environments, and ESs cannot be considered reliable and trustworthy. Thus, in edge computing-based systems, the integrity of  data  cached in ESs has to be  checked frequently  \cite{LI,LII}. 

According to the above explanation, in our considered setting, the app vendor has to frequently check the integrity of   replicas  cached in  ESs. However, the use of  existing integrity checking tools such as digital signatures and  provable data possession protocols   raises   new  challenges including high computational, storage, and I/O overhead (Challenges 2 and 3).

\textit{Challenge 2: Storage and I/O overhead.} Conventional integrity verification methods such as  digital signatures  take as input the entirety of  a  data file that its integrity is  to be verified \cite{shim}. Therefore,  if the app vendor employs these methods  or other approaches originating from digital signatures (e.g., \cite{LI}), it has to locally store  the  original data  files after caching the replicas in ECs. Obviously, in this case, the   storage and I/O overhead  incurred by the app vendor increases linearly with the amount  of  data  that it wants to  be shared with the users. It is clear that  if the app vendor could verify the integrity of the replicas without having the original data files, then it can remove the original files from its local memory to alleviate storage and I/O overhead  (instead, the app vendor can back all the original data files up onto a  cloud  server) \cite{HWang,Yuu}.

\textit{Challenge 3: Computational overhead.} Remote data inspection (RDI) protocols \cite{Yuu,dynamic,Blockchainnn,certificate-based}  offer promising solutions to Challenge 2. Using these approaches,  app vendors  can verify the integrity of replicas  without having the corresponding original data files.  Therefore,  RDI approaches can reduce the incurred storage and I/O overhead  satisfactorily. In these methods, at first, the app vendor tags  the replicas  to be given to ESs, and then it can remove the original data files from its local storage  systems. By using the assigned tags and a challenge-response protocol, the app vendor can verify the integrity of replicas. However, in  existing protocols, the tag generation process and the challenge-response protocol are highly costly   in terms of computational and communication overhead. This issue makes these approaches unsuitable for use in   EC environments,  as ESs and app vendors quite often are assumed to have limited  computational and communication  resources \cite{LI,alii1}. It is clear that providing an  RDI approach with  \textbf{batch verification of replicas} and \textbf{online/offline computation}  can alleviate the computational overhead.  Nevertheless,   new challenges such as  localization and repair of corrupted data files   may arise  if the app vendor employs batch verification methods for auditing the remote replicas (Challenge 4).

\textit{Challenge 4: Localization and repair of corrupted data.} Given a batch of data files,   the output of a batch  verification process usually is $\textbf{Yes}$ or $\textbf{No}$, where $\textbf{No}$ means that there is at least one corrupted data file in the  batch, and $\textbf{Yes}$ otherwise \cite{shim}.  However, these methods usually cannot detect the corrupted data file, whereas it is clear that the localization and repair of the corrupted replica are vital for the app vendor. On the other hand, if the app vendor employs RDI approaches and removes the original data files, it must be able to localize and repair the corrupted replicas cached in ESs without having the corresponding original data files.


\subsection{Main Contributions}

To address the described challenges, in this work, we
put forward a novel online/offline  data inspection (${\text{O}^2\text{DI}}$) scheme. Our main contributions are as follows:

\begin{enumerate}
	\item  \textit{Remote data inspection}: Our proposed approach enables the app vendor to check the integrity of replicas cached in ESs without needing the original data. Therefore, the app vendor can remove the original data files and their associated replicas after  caching them in the ESs. This feature addresses Challenges 1 and 2 efficiently.  
	\item \textit{Online/offline computation}:  In ${\text{O}^2\text{DI}}$,  almost all expensive computations related to the inspection process are performed in the offline mode, and the  online computations are extremely  fast and efficient. This feature   alleviates Challenge 3 effectively.
	
	\item \textit{Batch   verification}:  Given an arbitrary number of original data files and a set  of  ESs such that one replica of each data file has been cached  in all of the ESs, by  one-time execution of ${\text{O}^2\text{DI}}$, an app vendor can efficiently verify the integrity of the cached replicas  at once. This capability, together with the online/offline setting, can effectively address Challenge 3.
	
	\item \textit{Efficient localization and repair mechanism}:  Our ${\text{O}^2\text{DI}}$ approach enables the app vendor to easily detect which replicas in which ESs have been corrupted. Also, the app vendor can help the ESs to repair the corrupted replicas. Note that the app vendor can carry out both  localization and repair processes without needing the original data corresponding to the corrupted replicas. 
	
	\item \textit{Provable security}: We formalize the system model, threat model, and 	security definitions for our ${\text{O}^2\text{DI}}$ approach. Also, we 	prove that ${\text{O}^2\text{DI}}$ is secure under the hardness assumption  of the   bilinear Diffie-Hellman (BDH) problem.
\end{enumerate}

\subsection{Organization}
The rest of this work is organized as follows. Section \ref{rw}  overviews the  related work. We give   some required  preliminaries in Section \ref{Preliminaries}. In Section \ref{pfsm}, we  formalize our problem. Section \ref{47sd} presents our proposed ${\text{O}^2\text{DI}}$ approach in detail.
Sections \ref{kljhjk} and \ref{uhgjnba} present our security and performance analysis, respectively. 
Concluding remarks are provided  in Section \ref{c1c}. To make this work easy to follow and complete, we present the correctness analysis   in the appendix.

\section{Related Work}\label{rw}

Nowadays,  edge computing technology has  received tremendous attention in both academia and industry \cite{A12}. In EC environments, app vendors can hire  many ESs in geographically distributed locations to provide their users with low-latency   services such as computation, storage, data provision, etc  \cite{A5}.  In this regard, several  new challenges such as edge user allocation \cite{A13,A14},  edge application deployment \cite{A15,A16}, computation offloading \cite{A17,A18}, and edge server deployment \cite{A19,A20}  have been widely investigated. 
In particular, EC can facilitate  data    retrieval   process by providing caching systems with distributed ESs enabled  to render services to  users in specific geographical areas \cite{A21,A22}. Recently,  the edge data caching problem has attracted many researchers’ attention due to its great importance and various applications \cite{A23,A25}.

As a fundamental requirement, ensuring  the integrity of  data cached in  ESs is a big challenge for  caching systems \cite{LI}. Indeed, as we mentioned  before, the data cached in ESs is subject to a variety of threats and attacks \cite{LI,LII}. 
Digital signatures are known as  very effective cryptographic  tools to provide data integrity and authentication \cite{shim,A28}. Similar to other communication networks, in EC environments, digital signatures  have been widely used  to provide data integrity.  
Zhang et al.  \cite{Group}  designed  a group signature scheme  for blockchain-based  EC environments. Their scheme can effectively  counter   double-spend/long-range   and selfish mining attacks on blockchain-assisted EC systems. Also, by using group signatures and blockchain technology,  Wang \cite{Blockchain} designed  an  EC-based data storage system. In \cite{Proxy}, a non-interactive identity-based proxy re-signature scheme for IoT-assisted EC systems   has been put forward. The  scheme avoids costly pairing operations. Huang et al. \cite{NDN} proposed a  certificateless group signature scheme based on  EC in the NDN-IoT environment. Their scheme offers unforgeability,  anonymity, key escrow resilience, and traceability. Wang et al. \cite{Threshold} designed a privacy-preserving threshold ring signature scheme for use in EC environments. Li et al. proposed a multi-authority  attribute-based signature scheme for  EC-based systems \cite{SDABS}.

However, the verification processes in  digital signature schemes are usually  expensive in terms of computational overhead \cite{A26,A27}. Therefore,   as in  caching systems the number of replicas  cached in ESs  might be too high,   digital signatures seem not to be suitable choices for these applications. One might be tempted to think that the use of  aggregate signatures \cite{shim} instead of usual approaches can resolve this issue. An aggregate signature scheme enables a verifier to    condense a sequence of signed messages $(M_1,\sigma_1),\ldots,(M_N,\sigma_N)$  into a single aggregate signed message  $(M^*,\sigma^*)$ such that the validity of $(M^*,\sigma^*)$ implies the validity  of the  messages in the sequence \cite{shim}, and the invalidity of  $(M^*,\sigma^*)$ implies that there exists at least one  invalid message in the sequence. Therefore, such approaches can significantly reduce the computational costs associated with the verification process.  We refer  the readers to \cite{shim,A28} for more details   on this topic. 

Although aggregate signatures can lighten the computational overhead of data verification, the use of these approaches   may raise new challenges  for the localization of  corrupted data. Indeed, usual aggregate signatures  are only able to decide whether there is any  corruption in a set  of data files or not,  whereas in actual fact  they cannot determine  which of the data files have been corrupted \cite{shim,A28}.

To address this issue, Li et al. \cite{LI} proposed an efficient edge data inspection mechanism that employs an aggregate signature scheme as the main tool. Their designed approach can localize and repair the corrupted edge data efficiently.  In their proposed system,  given an original data file $M$ and edge servers $ES_1, \ldots,ES_N$, the app vendor makes replicas $M^{(1)},\ldots M^{(N)}$ of $M$ and caches $M^{(j)}$ in $ES_j$, $j=1,\ldots, N$. Also, the app vendor has to maintain the original data file $M$ in its local database, as it is needed for inspecting the replicas. We found their proposed system secure and efficient  in terms of computational complexity. However,  in their system, the app vendor  incurs prohibitive  storage and I/O costs, as it has to locally store the original data files.  

RDI protocols are  the most popular approaches to address the challenge of high storage  and I/O overhead in the data inspection process. Using these protocols, an app vendor can check the integrity of  a replica cached in an ES  without needing the associated original data file \cite{Yuu}. Therefore, these approaches can reduce storage and I/O overhead effectively.  The concept of RDI was proposed by Deswarte \cite{A28} for the first time.  In their work, the server has to hash the entirety of a data   file  to prove its integrity. However, it imposes  high  computational costs on the server when the volume of data is enormous. Filho et al. \cite{A29} designed  an RDI scheme that greatly improved the performance  of the inspection process.  Juels et al. \cite{A31} put forward the concept of Proof of Retrievability (PoR) and designed a concrete scheme. However, their  approach  can  only   support limited times of data inspection.   Ateniese et al. \cite{A30} independently put forward an analogous concept named Provable Data Possession (PDP).
Unlike PoR, PDP protocols   support    unlimited times of integrity checking and public auditing. The latter feature  is highly preferred, as  it enables users to  outsource the inspection task to a third-party auditor. Also, in the PDP protocol proposed in \cite{A32},  the communication and  I/O   costs have been decreased significantly by employing  random sampling and homomorphic authenticators.  Shacham and Waters \cite{A33} designed  a private and a public remote data auditing scheme by using   BLS signatures and pseudorandom functions.
Wang \cite{HWang} proposed a novel remote RDI scheme  for multi-cloud storage environments.  Their proposed scheme is  provably secure and provides    flexible and  efficient users' certificate management mechanism. 
Yu et al. put forward a privacy-preserving identity-based RDI scheme for cloud storage \cite{Yuu}.  Their proposed scheme leaks no information about the outsourced data to the auditor.
Shang et al. \cite{dynamic} designed  an identity-based dynamic RDI scheme. Their  proposed scheme supports dynamic data operations such as deletion, modification, and insertion.
Shu et al. \cite{Blockchainnn} designed  a  decentralized RDI scheme by using a decentralized blockchain network. They used a new concept of  decentralized autonomous organization to mitigate the threat of   malicious blockchain miners.
Li et al. \cite{certificate-based} proposed an efficient certificate-based RDI for cloud-assisted WBANs. In their scheme,  the computational overhead  incurred in the tag generation process  for a data block is constant, and it is not dependent upon the size of  data blocks.

However, to the best of our knowledge, in most existing   RDI schemes, the inspection procedure is conducted  as a challenge-response protocol with high  computation and communication overhead \cite{certificate-based}. Considering the limited resource of ESs, one concludes that the existing RDI approaches are not suitable  to use in EC environments. We believe that designing an RDI approach supporting    online/offline calculation can effectively lighten the overhead incurred by both app vendors and ESs.

In this work, we propose ${\text{O}^2\text{DI}}$ resolving all described challenges in this section. Table \ref{tab:ff} compares  ${\text{O}^2\text{DI}}$ with the described  state-of-the-art methods.

\begin{table}[t]  
	
	\caption{Comparison of Functionalities in various schemes.}\label{tab:ff}
	\centering
	\scalebox{0.99}{\begin{tabular}{|l|c|c|c|c|c|c|c|}
			\hline
			Schemes & $\mathbb{F}_1$ & $\mathbb{F}_2$ & $\mathbb{F}_3$ & $\mathbb{F}_4$ & $\mathbb{F}_5$& $\mathbb{F}_6$& $\mathbb{F}_7$\\
			\hline
			\hline
			Li et al. \cite{LI}  & &  \Checkmark &   &  \Checkmark &\Checkmark&  & \\
			\hline
			Wang \cite{HWang} & \Checkmark& &  &\Checkmark  &  &  &  \\
			\hline
			Yu et al.	\cite{Yuu} &\Checkmark &   &   &   & &  & \\
			\hline
			Shang et al.	\cite{dynamic} &\Checkmark &   &   &   &  &  & \Checkmark\\
			\hline
			
			Shu et al. \cite{Blockchainnn}  &\Checkmark &   &   & \Checkmark  &   &  & \\
			
			\hline
			Li et al. \cite{certificate-based} &\Checkmark &   &   &   &  &   &  \\
			\hline

			${\text{O}^2\text{DI}}$ &\Checkmark &\Checkmark  &\Checkmark  &\Checkmark  &\Checkmark & \Checkmark & \Checkmark\\
			\hline
			\hline
			\multicolumn{8}{c}{\textbf{Notes.} \scalebox{0.95}{$\mathbb{F}_1$: Inspection mechanism with low storage and I/O costs}} \\
			\multicolumn{8}{c}{\scalebox{0.97}{\,\,overhead; $\mathbb{F}_2$: Batch data verification; $\mathbb{F}_3$: Offline-online setting;}}\\
			\multicolumn{8}{c}{\scalebox{0.96}{\,\,$\mathbb{F}_4$: Distributed  caching system; $\mathbb{F}_5$: Corrupted data Localization}}\\
			\multicolumn{8}{c}{\scalebox{0.97}{and repair; $\mathbb{F}_6$: Lightweight tagging process; $\mathbb{F}_7$: Dynamic data}}\\
			\multicolumn{8}{c}{{ operation.}\,\,\,\,\,\,\,\,\,\,\,\,\,\,\,\,\,\,\,\,\,\,\,\,\,\,\,\,\,\,\,\,\,\,\,\,\,\,\,\,\,\,\,\,\,\,\,\,\,\,\,\,\,\,\,\,\,\,\,\,\,\,\,\,\,\,\,\,\,\,\,\,\,\,\,\,\,\,\,\,\,\,\,\,\,\,\,\,\,\,\,\,\,\,\,\,\,\,\,\,\,\,\,\,\,\,\,\,\,\,\,\,\,\,\,\,\,\,\,\,\,\,\,\,\,\,\,}\\
	\end{tabular}}
\end{table}

\section{Preliminaries} \label{Preliminaries}
Assume that  $\mathcal{A}$ is a probabilistic polynomial-time (PPT) algorithm. Let us assume that $O\leftarrow \mathcal{A}$ denotes the execution of  $\mathcal{A}$ on input $I$  that assigns the output to $O$.  In this section, we briefly present some preliminaries required for the rest of this paper. 
\subsection{Bilinear map}\label{jhjhhjh}
Let  $G_1$ and $G_2$ be two cyclic groups  of a prime order $q$. We say that a function $\hat{e}:G_1 \times G_1 \to G_2$ is  bilinear  if the following conditions hold: (1) 
\textbf{Bilinearity}--For any  $a, b \in \mathbb{Z}_p$ and $g \in G_1$, we have  $\hat{e}(g^{a},g^{b})=\hat{e}(g^{b},g^{a})=\hat{e}(g,g)^{ab} $;
(2) \textbf{Non-degeneracy}-- $\hat{e}(g,g)\ne 1$, for at least one $g\in G_1$;
(3) \textbf{Computability}--There exists a PPT algorithm that computes $\hat{e}(g,h)$, for any $g,h \in G_1$.

\subsection{Hardness  assumptions}
Our security analysis given in Section \ref{kljhjk}  is conducted according to the hardness assumption of  the  bilinear Diffie-Hellman (BDH) and Diffie-Hellman (DH)  problems described as follows:
\subsubsection{Bilinear Diffie-Hellman (BDH) problem}
Assume that $\mathcal{G}$ is  a PPT algorithm  that, on input a security parameter $n$, outputs a tuple   $(q, G_1, G_2, \hat{e})$, where  $p$, $G_1$, $G_2$, and $\hat{e}$ are the same as in Subsection \ref{jhjhhjh}.
We state  that the  decisional bilinear Diffie-Hellman (DBDH) assumption holds for $\mathcal{G}$ if for any  tuple {$(n, q, G_1, G_2, \hat{e}, g, g^ {\alpha}, g^{\beta}, g^{\gamma})$} and  all PPT adversaries $\mathcal{A}$ we have $\Pr[\mathcal{A}(n, q, G_1, G_2, \hat{e}, g, g^ {\alpha}, g^{\beta}, g^{\gamma})=\hat{e}(g,g)^{\alpha\beta\gamma}]$   is a negligible function in $n$, where $(n,q,G_1,G_2,\hat{e})\leftarrow\mathcal{G}(1^n)$ and $g \in G_1$, $\alpha, \beta, \gamma \in \mathbb{Z}_q$ are selected uniformly at random.

\subsubsection{Diffie-Hellman (DH) problem}
Consider a PPT algorithm  $\mathcal{G}'$ that, on input a security parameter $n$, returns a  group $G$ of a prime order $q$. 
We say  that the  Diffie-Hellman (DH) problem is hard relative to $\mathcal{G}$  if for any $(n, p, G, g, g^{x},g^y)$  the advantage of   any PPT adversary $\mathcal{A}$ in the calculation  of $g^{xy}$  is a negligible function in  $n$, where   {$(n,p, G)\leftarrow \mathcal{G}'(1^n)$}, and $g\in G$, $x,y\in \mathbb{Z}_q$ are selected uniformly at random.
%
\subsubsection{Identity-based  signatures (IBS)}
An IBS scheme ${\Pi}$ is a tuple of  four PPT algorithms $({\textbf{Gen}},{\textbf{KeyGen}},{\textbf{Sign}},{\textbf{Vrfy}})$ such that:
\begin{itemize}
	\item 
	${\textbf{Gen}}(1^{n})$: Given a security parameter $n$, it generates  public parameters $params_s$ and the master secret key ${MSK}_s$.
	\item ${\textbf{KeyGen}}(params_s,{MSK}_s,ID)$: On input the public parameters $params_s$, the master secret key ${MSK}_s$, and an identifier $ID$, this algorithm returns a secret key $sk$.
	\item $\textbf{Sign}(params_s,ID,sk,M)$: This algorithm takes as input the public parameters ${PK}$, an identifier $ID$,  a secret key ${sk}$, and a message $M$,  it returns a signature $\sigma$.
	\item ${\textbf{Vrfy}} ({PK},\sigma,ID,M)$: On input the public parameters $params_s$, a signature $\sigma$, an identifier  $ID$, and a  message $M$, this algorithm outputs a bit $b\in \{0,1\}$, where $b = 1$ means that  $\sigma$ is a valid signature associated with $M$ and $ID$, and
	$b = 0$ means that it is invalid. 
\end{itemize}
In addition to the described primitives,  our ${\text{O}^2\text{DI}}$ approach benefits from famous cryptographic primitives such as  hash  and pseudorandom  functions. We refer the reader to \cite{modernn} for further details on  these matters.
\section{Problem Formulation}\label{pfsm}
To make it easier to follow our work, in this section, we describe our  system and threat models, then we present our security requirements.

\subsection{System Model} 
In our intended  EC system, there exists an app vendor (AV) hiring a number of edge servers (ESs) to cache its data and provide the data for  users within the ESs' coverage areas \cite{LI}. Suppose that there exist $N$ hired edge servers $ES_1,\ldots,ES_N$ in our system.   Let $M$ denote an original data file and $M^{(1)},\ldots, M^{(N)}$ be the  edge data replicas of $M$, where  $M^{(j)}$ has been cached in  $ES_j$, $j=1,\ldots N$.
As shown in Figure \ref{System}, after caching the replicas in ESs, the AV should  continuously interact  with  ESs to i) check the integrity of the replicas; ii) localize the corrupted  replicas, and iii) repair the corrupted data  blocks. In our proposed system, by using our designed ${\text{O}^2\text{DI}}$, the AV can carry out the tasks efficiently. It should be notified that one of the features  making our approach different from similar techniques (e.g., \cite{LI}) is that in our ${\text{O}^2\text{DI}}$ the AV does not need the original data file after caching the replicas. Indeed, it can conduct the inspection, localization, and repair processes without having the original data. This feature decreases the memory  and I/O overhead on the AV side significantly. 

\subsection{Threat Model and Design Goals} \label{sr}
In our considered system, the main security concerns originate from   \textbf{edge data corruption} and \textbf{forging  data integrity proof}  which are briefly described below: 
\begin{itemize}
	\item 
	\textbf{Edge data corruption}: In this work, ESs are not assumed to be trustworthy in terms of reliability and validity. Indeed, because of their limited computational resources and   geographical distribution, ESs are subject to a variety of  intentional or unintentional attacks \cite{A5,LI,A11,LII,A14}. Therefore,  the edge data replicas cached in   ESs may  be tampered with or destroyed.
	\item 
	\textbf{Forging  data integrity proof}: Similar to cloud service providers \cite{Yuu,Alip},   edge infrastructure providers (EIP) may try to cover up  data loss incidents   in order to  gain benefits or  circumvent the compensation. Indeed, when some ESs lose their  data, they may try to provide some integrity proofs deceiving the AV about the accuracy of the cached data.
\end{itemize}

To tackle the threats described above,  our  ${\text{O}^2\text{DI}}$ scheme should provide the following features:

\begin{itemize}
	\item 
	\textbf{Unforgeability}: This requirement states	that the ESs, whose data has been corrupted,  must not be able to   provide a valid integrity proof. 
	\item 
	\textbf{High efficiency}: Our proposed approach should be highly efficient for both the AV and ESs. The AV should be able to quickly tag the replicas  and     remotely inspect them  with minimal computational costs after caching them in ESs.  Moreover, ESs should be able to prove the replicas' integrity  with low computation and communication overhead. 
	\item 
	\textbf{Effectiveness}:  Our proposed approach should  provide the following services: 1) It should enable   the  AV to easily detect any edge data corruption; 2) the AV should be able to accurately  localize the corrupted replicas; 3)  The ESs should be able to repair the  corrupted replicas with  the minimum  assistance of the AV.  
\end{itemize}

\subsection{Overview of ${\text{O}^2\text{DI}}$ Mechanism }  
\begin{figure}[h]
	\centering
	\includegraphics[width=3.5in]{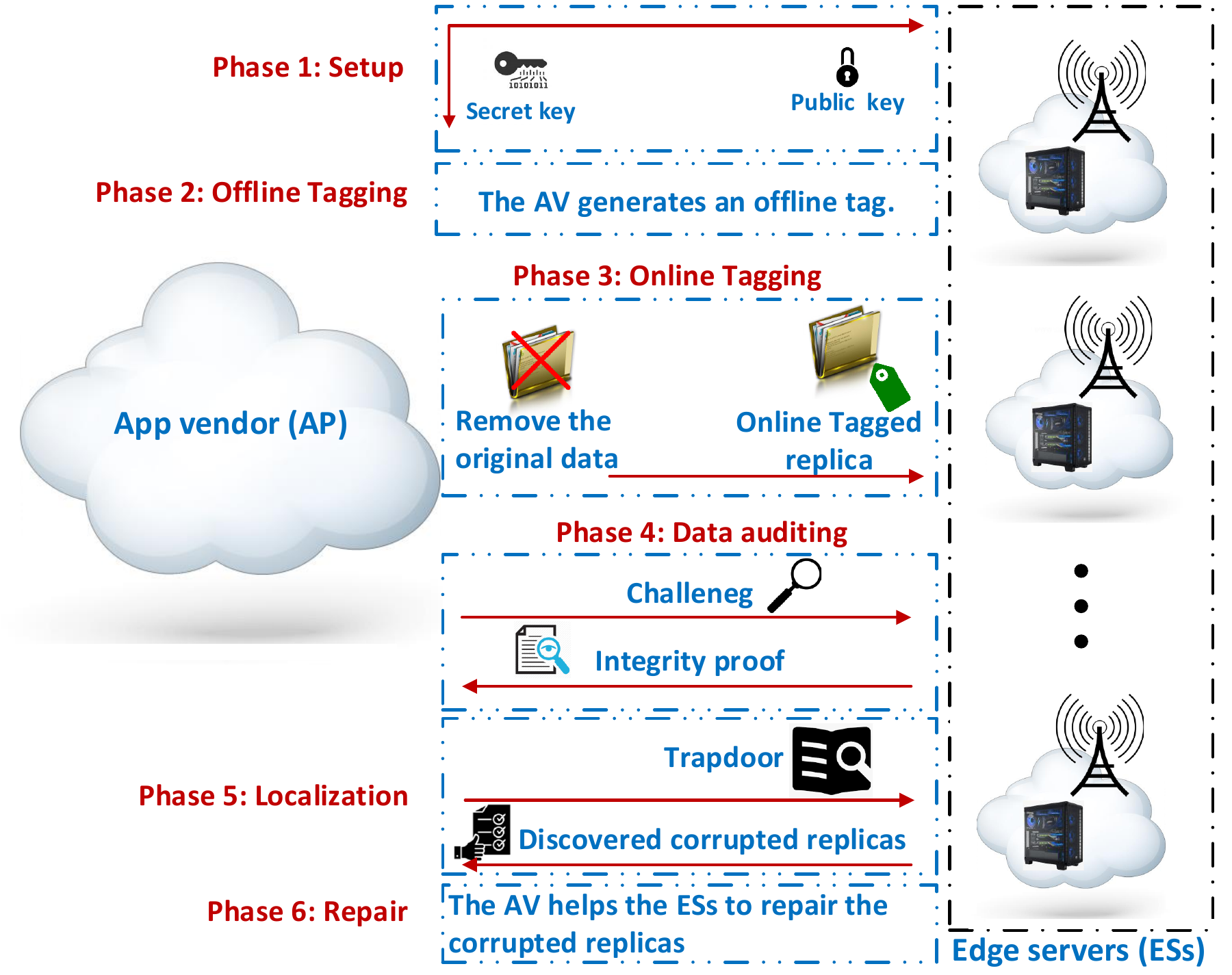} 
	\caption{Workflow of our considered caching system. }
	\label{System}
\end{figure}

Although in practice ESs are distributed in different geographical regions of the world,  our ${\text{O}^2\text{DI}}$ approach can carry out the edge data   inspection and other related processes such as localization and repair of the corrupted data in an aggregate manner. Specifically, our approach consists of six basic phases \textbf{Setup}, \textbf{Offline tagging}, \textbf{Online tagging}, \textbf{Data inspection}, \textbf{Localization}, and 	\textbf{Repair} which are briefly described below (see Figure \ref{System}).

\begin{itemize}
	\item 
	\textbf{Setup}: The AV initiates public parameters and its own secret keys that will be utilized in the subsequent phases. 
	\item 
	\textbf{Offline tagging}: To make the remote data inspection possible, the AV should tag the data files  before caching them in ESs. In this phase,  all expensive computational operations required to tag the data files are performed  in the offline mode.  
	\item 
	\textbf{Online tagging}: Given a data file to be shared with users of the system,  in this phase, the AV first makes several replicas of the data file. Then,  by using the offline tags generated in the previous phase and performing highly efficient operations, the AV tags the replicas. One replica, along with its associated tag,  is given   to  any ES.  
	\item 
	\textbf{Data inspection}: In this phase, by performing a challenge-response protocol, the AV checks the integrity of   the  replicas cached in   ESs. It should be noticed that the inspection approach used in this phase supports batch verification. Indeed, given a batch of data files $\{M_s\}_{s\in S}$ and its associated replicas $\{\{M_{s}^{(j)}\}_{j=1}^{N}\}_{s\in s}$, where $\{M_{s}^{(j)}\}_{j=1}^{N}$ is the set of replicas of $M_s$, the AV can check the integrity of $\{\{M_{s}^{(j)}\}_{j=1}^{N}\}_{s\in S}$ by  only  one-time execution of the challenge-response protocol such that the computational cost of the batch inspection process  is almost equal to the cost of  a single replica inspection. 
	\item 
	\textbf{Localization}: Although the batch verification technique can reduce   computational costs, it may make the detection of  corrupted replicas difficult. Indeed, given an original data file $M$ and a batch of its replicas $\{M^{(j)}\}_{j=1}^N$ cached in some ESs,  if the result of our inspection test  shows that some replicas have been corrupted, the AV needs to detect the  ESs whose replicas are corrupted. In this phase,  by running an interactive protocol, the AV determines which replicas  in which  ESs need to be repaired.
	
	\item 
	\textbf{Repair}: After localizing the corrupted replicas, the AV helps the ESs  to repair their replicas. 
\end{itemize}

\begin{table}[h]   
	\caption{Notations.} \label{tab:fa}
	\centering
	\scalebox{0.9}{	\begin{tabular}{|l|c|}
			\hline
			Notations & Descriptions\\
			\hline
			\hline
			$n$ & 	\myalign{l}{Security parameter of the system}\vline \\
			\hline
			$params$ &  	\myalign{l}{Public parameters of the system}\vline\\
			\hline
			$MSK_{av}$ &   		\myalign{l}{Master secret key of the AV}\vline\\
			\hline
			$SK_{av}$& 	\myalign{l}{Secret key of the AV}\vline\\
			\hline
			$PK_{av}$ &  	\myalign{l}{Public key  of the AV}\vline\\\hline
			$ID_{av}$ & 	\myalign{l}{Identifier of the AV}\vline\\
			\hline
			$Off\text{-}Tag_{av}$ & 		\myalign{l}{Offline tag generated by the AV}\vline\\
			\hline		
			$POff\text{-}Tag_{av}$ & 	\myalign{l}{Public part of the offline tag generated by the AV}\vline\\
			\hline		
			$SOff\text{-}Tag_{av}$& 		\myalign{l}{Secret part of the offline tag generated by the AV}\vline\\
			\hline		
			$ES_j$& 		\myalign{l}{The j-th edge server in our designed system}\vline\\
			\hline
			$N$& 	\myalign{l}{Number of edge servers in our designed system}\vline\\
			\hline	
			$M$& 	\myalign{l}{An original data file}\vline\\
			\hline	
			$M^{(j)}$&  	\myalign{l}{A replica of $M$ cached in the edge server $ES_j$}\vline\\
			\hline	
			$\{M_s\}_{s\in S}$&  		\myalign{l}{Set of original data files}\vline\\
			\hline
			$M_s^{(j)}$& 		\myalign{l}{A replica of $M_s$ cached in the edge server $ES_j$}\vline\\
			\hline		
			$On\text{-}Tag^{(j)}$   & 	\myalign{l}{Online tag corresponding to the $j$-th replica of a data file}\vline\\
			\hline
			$C$& 	\myalign{l}{A challenge}\vline\\
			\hline
			$k$&  	\myalign{l}{A challenged secret }\vline\\
			\hline
			$\tilde{k}$& 	\myalign{l}{A secret key}\vline\\
			\hline
			$id_{M^{(j)}}$& 	\myalign{l}{Identifier of a replica $M^{(j)}$}\vline\\
			\hline
			${C^{(j)}}$& 		\myalign{l}{A challenge given to the edge server $ES_j$}\vline\\
			\hline
			${P^{(j)}}$& 	\myalign{l}{An integrity proof generated by the edge server $ES_j$}\vline\\
			\hline
			$S$& 		\myalign{l}{The index set corresponding to original data files}\vline\\
			\hline
			$M^{Agg}$& 	\myalign{l}{An aggregate data file}\vline\\
			\hline
			$M^{(Agg,j)}$& \myalign{l}{An aggregate data file generated by the edge server $ES_j$}\vline\\
			\hline
			$On\text{-}Tag^{Agg}$& 	\myalign{l}{An aggregate online tag}\vline\\
			\hline
			$On\text{-}Tag^{(Agg,j)}$& 	\myalign{l}{An aggregate online tag generated by the edge server $ES_j$}\vline\\
			\hline
			$P^{Agg}$& 	\myalign{l}{An aggregate integrity proof}\vline\\
			\hline
			$P^{(Agg,j)}$& 	\myalign{l}{An aggregate integrity proof generated by  $ES_j$}\vline\\
			\hline
			$K^{Agg}$& 	\myalign{l}{An aggregate challenged secret}\vline\\
			\hline
			$C^{Agg}$& 	\myalign{l}{An aggregate challenge}\vline\\
			\hline
			$PreChall$& 	\myalign{l}{Precomputed challenges}\vline\\
			\hline
			$Trap_S^{(j)}$& 	\myalign{l}{A trapdoor given to the edge server $ES_j$}\vline\\
			\hline
			$S'_j$& 	\myalign{l}{The index set corresponding to  corrupted replicas in  $ES_j$}\vline\\
			\hline
	\end{tabular}}

\end{table}


\section{ Proposed ${\text{O}^2\text{DI}}$ approach}\label{47sd}
In this section, we present our proposed ${\text{O}^2\text{DI}}$ approach in detail. Figure \ref{ajnaa}  describes the algorithms employed in ${\text{O}^2\text{DI}}$. Table \ref{tab:fa} describes the notations used in this section. As we mentioned before, our ${\text{O}^2\text{DI}}$ consists of six phases \textbf{Setup}, \textbf{Offline Tagging}, \textbf{Online Tagging}, \textbf{Data inspection}, \textbf{Localization}, and \textbf{Repair} which are exhaustively described as follows.

\subsection{Setup}
In this phase, the AV runs  the algorithm $(params, MSK_{av})\leftarrow\mathbf{Setup}(1^n)$ to generate   public parameters $params$ and its own master secret key $MSK_{av}$. It makes $params$ publicly available and keeps $MSK_{av}$ completely  confidential. 

\subsection{Offline Tagging}
Prior to  caching the replicas in ESs, the AV should  tag them to make the remote inspection  possible. Generally, the tagging process incurs prohibitive computational costs. However,   in our ${\text{O}^2\text{DI}}$ approach,  the AV can perform the expensive operations only  once in the offline mode, and very efficient operations are left   to the online tagging stage. In this phase, the AV generates an offline tag by running  $(SK_{av},PK_{av})\leftarrow\textbf{Extact}(params,MSK_{av},ID_{av})$ and $Off\text{-}Tag_{av}\leftarrow\textbf{Offline-Tag}(params,PK_{av},SK_{av})$.  The offline tag $Off\text{-}Tag_{av}$ is divided into two parts pubic  tag  $POff\text{-}Tag_{av}$ and secret  tag $SOff\text{-}Tag_{av}$. As expected, the public part is made publicly available and the secret tag is kept confidential. 

\subsection{Online Tagging} 
Given an original  data file $M$, assume that the AV wants to cache $M$ in edge servers $ES_1,\ldots, ES_N$. To this end, it first makes replicas  $M^{(1)},\ldots,M^{(N)}$ of $M$ and selects a  unique identifier  $id_{M^{(j)}}$ for  $M^{(j)}$,   $j=1,\ldots,N$. Then,   using the offline tag $Off\text{-}Tag_{av}$ obtained in the previous phase,  the AV executes  \scalebox{0.98}{$On\text{-}Tag^{(j)}\leftarrow \textbf{Online-Tag}(params,id_{M^{(j)}},Off\text{-}Tag_{av},M^{(j)},$}\\$SK_{av})$ to generate an online tag for  $M^{(j)}$,  $j=1,\ldots,N$. It stores  $id_{M^{(j)}}$ and sends $(M^{(j)},On\text{-}Tag^{(j)})$ to $ES_j$. 

It is worth noting   after caching the replicas $M^{(1)},\ldots,M^{(N)}$ in the ESs, the AV does not need the replicas, and even the original data file $M$, for inspecting the cached replicas. Indeed,   the AV can remove $M$ and $M^{(1)},\ldots,M^{(j)}$ from its local database and  store just $M$ on  a cloud server.

\subsection{Data inspection}\label{2154}
In this phase, the AV inspects the integrity of the replicas cached in  ESs.   This phase can be performed in four different methods described below:

\textbf{ {Method 1: One data file and one ES}}.
Given an original  data file $M$ and its replica $M^{(j)}$ cached in the edge server $ES_j$,  this method enables the AV to check the integrity of $M^{(j)}$ without having   $M$ and $M^{(j)}$.

In this method, the AV first considers  the identifier $id_{M^{(j)}}$, associated with the replica  to be inspected, and executes 	$(C,k)\leftarrow\textbf{ChallGen}_1(params,PK_{av})$ to generate a  challenge $C$ and a secret $k$. It keeps the secret $k$ confidential and    returns    $(C,id_{M^{(j)}})$ to $ES_j$. When $ES_j$ receives the  challenge, it  calls the replica $M^{(j)}$ and  the tag $On\text{-}Tag^{(j)}$ associated with  $id_{M^{(j)}}$. Then, it runs \scalebox{0.94}{$P^{(j)}\leftarrow\textbf{GenProof}(params,C,On\text{-}Tag^{(j)},PK_{av},M^{(j)},ID_{av})$} to generate a proof of integrity $P^{(j)}$. The proof $P^{(j)}$ is given to the AV. By running \scalebox{0.99}{$b=\textbf{CheckProof}_1(params, POff\text{-}Tag,C,k,P^{(j)})$}, the AV  checks the integrity proof, where $b\in \{0,1\}$.  The AV approves the proof and thus the integrity of  ${M^{(j)}}$ only if $b=1$.

\textbf{ {Method 2: Multiple data files and one ES}}.
Given a batch of original data files $\{M_{s}\}_{s\in S}$ and its associated  replicas $\{M_{s}^{(j)}\}_{s\in S}$ cached in  $ES_j$, where $|S|>1$,  using this method, the AV can efficiently  verify the integrity of $\{M_{s}^{(j)}\}_{s\in S}$ at once without knowing $\{M_{s}\}_{s\in S}$ and   $\{M_{s}^{(j)}\}_{s\in S}$.

Considering  $\{id_{M^{(j)}_s}\}_{s\in S}$, the AV first executes   the algorithm $(C,k)\leftarrow\textbf{ChallGen}_1(params,PK_{av})$ to generate a challenge $C$ and its associated secret $k$. It also selects a random key $\tilde{k}$ and  sends a challenge $(C,\{(id_{M^{(j)}_s}\}_{s\in S}, \tilde{k})$ to $ES_j$. When  $ES_j$ receives the challenge, it first calls the associated tagged replicas $(\{M^{(j)}_s,On\text{-}Tag^{(j)}_{s}\}_{s\in S})$ and aggregates them by running  
$(M^{Agg},On\text{-}Tag^{Agg})\leftarrow\textbf{AggData}(params,\tilde{k},\{On\text{-}Tag_{s}^{(j)}\}_{s\in S},\{M_s^{(j)}\}_{s\in S})$. Then, it executes $P^{Agg}\leftarrow\textbf{GenProof}(params,C,On\text{-}Tag^{Agg}, PK_{av}, M^{Agg},ID_{av})$ to generate the requested proof. Finally, to check the soundness of the proof, the AV executes  $b=\textbf{CheckProof}_2(params, Off\text{-}Tag,C,k,\tilde{k},P^{Agg}, \{id_{M_s}\}_{s\in S})$. The data integrity is approved if and only if $b=1$.

\textbf{ {Method 3: One data file and Multiple ESs}}.
Consider an original data file $M$ and its associated    replicas $\{M^{(j)}\}_{j=1}^{N}$ such that $N>1$ and $M^{(j)}$ has been cached in  $ES_j$, $j=1,\ldots,N$. This method enables the AV to verify the integrity of $\{M^{(j)}\}_{j=1}^{N}$ at once  without needing  $M$ and   $\{M^{(j)}\}_{j=1}^{N}$.

In this method, before starting the inspection process, the AV can perform most  operations related to the challenge generation in the offline mode. To this end,  it runs $PreChall\leftarrow\textbf{Offline\text{-}Challenge}(params,1^m,PK_{av})$, for a security parameter $m$. Then, using the set of values  $PreChall$, it can generate polynomially many challenges with minimal computational cost.
When  the AV  wants to inspect the replicas associated with $\{id_{M^{(j)}}\}_{j=1}^{N}$, it runs   	$(C_{Agg},K_{Agg})\leftarrow\textbf{ChallGen}_2(params,PreChall,N)$ to generate a bath of challenges $C_{Agg}=\{C^{(j)}\}_{j=1}^{N}$ and the associated secret $K_{Agg}$ efficiently. Then, for $j=1,\ldots,N$, the challenge $C^{(j)}$ is given to $ES_j$, and  $K_{Agg}$ is kept confidential. Afterwards,  $ES_j$  executes  
\scalebox{0.96}{$P^{(j)}\leftarrow\textbf{GenProof}(params,C,On\text{-}Tag^{(j)},PK_{av},$}\\$M^{(j)},ID_{av})$ and gives the generated proof $P^{(j)}$ to the AV. Finally, by running   	\scalebox{1}{$b=\textbf{CheckProof}_3(params, Off\text{-}Tag,C_{Agg},K_{Agg},$}\\{$\{P^{(j)}\}_{j=1}^N)$}, the AV checks whether the replicas are integral or not. It confirms the replicas' integrity if and only if $b=1$.

\textbf{ {Method 4: Multiple data files and Multiple ESs}}.
Consider a batch of  original data files   $\{M_{s}\}_{s\in S}$  and its associated    replicas  $\{\{M_s^{(j)}\}_{j=1}^{N}\}_{s\in S}$, where $|S|>1$, $N>1$, and $M_{s}^{(j)}$ is the replica of $M_s$ that it is cached in the edge server $ES_j$. By using this method, the AV can check the integrity of   $\{\{M_s^{(j)}\}_{j=1}^{N}\}_{s\in S}$ at once efficiently. The same as before,   the AV does not need  $\{M_{s}\}_{s\in S}$  and $\{\{M_s^{(j)}\}_{j=1}^{N}\}_{s\in S}$ for inspecting  the replicas.

Similar to Method 3,  the AV can carry out  expensive  computations associated with the challenge generation in the offline mode. To this end, it  first  selects a security parameter $m$ and runs \scalebox{1}{$PreChall\leftarrow\textbf{Offline\text{-}Challenge}(params,1^m,PK_{av})$}. Then, to check the integrity of replicas corresponding to the identifiers $\{\{id_{M_s^{(j)}}\}_{j=1}^{N}\}_{s\in S}$, the AV executes  $(C_{Agg},K_{Agg})\leftarrow\textbf{ChallGen}_2(params,PreChall,N)$ to obtain  challenges $C_{Agg}=\{C^{(j)}\}_{j=1}^{N}$ and a set of secrets  $K_{Agg}$. Also, it selects a  secret key $\tilde{k}$   and gives $(\tilde{k},C^{(j)})$ to $ES_j$,    $j=1,\ldots,N$.   $ES_j$ executes the algorithm \scalebox{1}{$(M^{(Agg,j)},On\text{-}Tag^{(Agg,j)})\leftarrow\textbf{AggData}(params,\tilde{k},$}\\\scalebox{0.92}{$\{On\text{-}Tag_{s}^{(j)}\}_{s\in S},\{M_s^{(j)}\}_{s\in S})$}
and  $P^{(Agg,j)}\leftarrow\textbf{GenProof}(params,On\text{-}Tag^{(Agg,j)},C_{Agg},PK_{av},M^{(Agg,j)},ID_{av})$
to generate a proof of possession \scalebox{1}{$P^{(Agg,j)}$}, $j=1,\ldots,N$. When the AV receives  \scalebox{1}{$\{P^{(Agg,j)}\}_{j=1}^{N}$}, it can check the integrity of  replicas  $\{\{M_s^{(j)}\}_{j=1}^{N}\}_{s\in S}$ by running  $b=\textbf{CheckProof}_3(params,$
\scalebox{0.98}{$ Off\text{-}Tag,C_{Agg},K_{Agg},\{P^{(Agg,j)}\}_{j=1}^N)$}, where $b\in \{0,1\}$.  The AV approves the replicas' integrity if and only if $b=1$.  
\subsection{Localization}\label{45444}
The localization approach can be described independently of the inspection method utilized in the previous phase.  


Consider replicas $\{\{M^{(j)}_s\}_{j=1}^{N}\}_{s\in S}$, where $N\ge 1$ and $|S|\ge 1$. Assume that the AV has realized that some of the replicas have been corrupted. In this case, to    identify the corrupted replicas, it runs \scalebox{0.93}{$Trap_{S}^{(j)}\leftarrow\textbf{Loc\text{-}Trap}(params,\{id_{M_s^{(j)}}\},C,k,SOff\text{-}Tag)$} to generate a trapdoor  $Trap_{S}^{(j)}$,  $j=1,\ldots,N$. It sends $Trap_{S}^{(j)}$ to  $ES_j$, for $j=1,\ldots,N$. Afterward,  by executing $S'_{j}\leftarrow$ \scalebox{0.97}{$\textbf{Find\text{-}Corrupted}(params, POff\text{-}Tag,\{M_{s}^{(j)}\}_{s\in S},C,Trap_S^{(j)})$},   $ES_j$ can easily discover   an index set $S'_{j}\subseteq S$  corresponding to  its corrupted replicas.   $S'_{j}$ is given to the AV.  

\subsection{Repair}
After localizing the corrupted replicas, the AV refers the ESs whose replicas have been corrupted to ESs with correct replicas. In this way, the corrupted data  can be easily recovered. More precisely, given the index set $\{S'_j\}_{j=1}^N$ obtained through  the localization process, for each $j^*\in \{1,\ldots,N\}$ and $s\in S'_{j^*}$ the AV  finds a $j\ne j^*$ such that $s\notin S'_j$ and refers $ES_{j^*}$ to $ES_j$ to be helped to repair the corrupted  replica $M_s^{(j^*)}$.



\begin{figure*}
	
	\begin{textbox}{box:somelabel}{}
		%{
		\begin{myframe} \label{ajnaaa}
			{\small
				\textbf{Setup}$(1^n)$: On input a security parameter $n$, it first executes $(n,q,G_1,G_2,\hat{e})\leftarrow\mathcal{G}(1^n)$ and selects   $g\in G_1$ and $x\in \mathbb{Z}_q$ uniformly at random. Then, it selects two secure hash functions $H:\{0,1\}^*\to G_1$ and $H':\{0,1\}^*\to \mathbb{Z}_q$, a pseudorandom function $F:\mathbb{Z}_q \times \mathbb{Z}_q \to \mathbb{Z}_q$,  and a secure IBS scheme $\Pi_s=(\textbf{Gen},\textbf{KeyGen},\textbf{Sign},\textbf{Vrfy})$  and an identifier $ID^*$. It executes $(MSK_s,params_s)\leftarrow \textbf{Gen}(1^n)$ and  $sk^*\leftarrow\textbf{KeyGen}(params_s,MSK_s,ID^*)$ and computes $pk=g^{x}$. Also, it considers a natural number $\ell$ as the number of blocks in each data file. Finally, this algorithm outputs a  secret key $MSK=(sk^*,x,MSK_s)$ and public parameters $params=(n,q,g,G_1,G_2,\hat{e},H,H',F,\Pi_s,ID^*,params_s,pk,\ell)$.
\vspace{0.15cm}
				\\
				
				\noindent$\textbf{Extact}(params,MSK,ID)$: It selects  $ \lambda\in \mathbb{Z}_q$ uniformly at random and computes   $\Lambda=g^{\lambda}$,  $sk\leftarrow\textbf{KeyGen}(Params_s,MSK_s,ID)$, and $\sigma\leftarrow\textbf{Sign}(params_s,sk^*,\Lambda)$.  It returns $SK=(\lambda, sk)$ as the secret key  and  $PK=(\Lambda,\sigma)$ as the public key.
\vspace{0.15cm}
				\\
				
				$\textbf{Offline-Tag}(params,PK,SK)$: On input a public key $PK=(\Lambda,\sigma)$ and secret key  $SK=(\lambda, sk)$, this algorithm selects $y\in \mathbb{Z}_q$ uniformly at random and computes $T_1=g^{y}$ and $T_2=pk^{y}$. It runs $\sigma'\leftarrow\textbf{Sign}(params_s,sk,T_1)$ and considers a public offline tag $POff\text{-}Tag=(T_1,\sigma',PK)$ and a secret offline tag $SOff\text{-}Tag=(T_2,y)$. It outputs $Off\text{-}Tag=(POff\text{-}Tag,SOff\text{-}Tag)$.
\vspace{0.15cm}
				\\
				
				$\textbf{Online-Tag}(params,id_M,Off\text{-}Tag,M,SK)$: Given an identifier $id_M$, an  offline  tag $Off\text{-}Tag=(T_1,T_2,y,\sigma',PK)$,  a data file $M=(m_1,m_2,\ldots,m_\ell)$,  and the secret key  $SK=(\lambda,sk)$, it    calculates ${t_i} = {\lambda}{m_i} +yH'(i{d_M}||T_2||i)$, for  $i=1,\ldots,\ell$, and returns an  online tag $On\text{-}Tag=(id_M,\{t_i\}_{i=1}^\ell)$.
\vspace{0.15cm}
				\\
				
				$\textbf{PreCheck}(params,POff\text{-}Tag)$: Given $POff\text{-}Tag=(T_1,\sigma',PK)$, where $PK=(\Lambda,\sigma)$,  it verifies the signatures $\sigma$ and $\sigma'$. If  both  of them  are  valid, then it returns $1$, and $0$ otherwise. 
\vspace{0.15cm}
				\\
				
				$\textbf{ChallGen}_1(params,PK)$: On input $PK=(\Lambda,\sigma)$, it selects $k\in \mathbb{Z}_p$ uniformly at random and computes $c_1=g^k$ and $c_2=\Lambda^{k}$. Then, it  considers a uniform subset $I\subseteq \{1,\ldots, \ell\}$  and  a random element $v_i \in \mathbb{Z}_p$,  for each $i\in I$. It outputs a challenge $C=(c_1,c_2,\{(i,v_i)\}_{i\in I})$ and its associated secret $k$.
\vspace{0.15cm}
				\\
				
				$\textbf{GenProof}(params,C,On\text{-}Tag,PK,M,ID)$: Given $C=(c_1,c_2,\{(i,v_i)\}_{i\in I})$, $On\text{-}Tag=(id_M,\{t_i\}_{i=1}^\ell)$, $PK=(\Lambda,\sigma)$,  a message $M=(m_1,m_2,\ldots,m_\ell)$, and an identifier $ID$, it first checks whether $\hat{e}(c_2,g)=\hat{e}(c_1,\Lambda)$. If not, it aborts. Otherwise,  it  computes $\mu  = \sum\limits_{i \in I} {{v_i}{m_i}}$,  $P_1=\hat{e}(H{(ID)^{\sum\nolimits_{i \in I} {{v_i}{t_i}} }} ,{c_1})$, and $P_2=\hat{e}(H(ID),{c_2}^{-\mu})$. Finally, it outputs a proof  $P=(P_1,P_2)$.
\vspace{0.15cm}
				\\
				
				$\textbf{CheckProof}_1(params, POff\text{-}Tag,C,k,P)$: On input   $POff\text{-}Tag=(T_1,\sigma',PK)$,  a challenge $C=(c_1,c_2,\{(i,v_i)\}_{i\in I})$, a secret $k$, and a proof $P=(P_1,P_2)$,  this algorithm returns 1 if and only if the following equation holds: 
				\begin{align}\label{ajakja}
					P_1P_2=\hat{e}(H(ID),T_1)^{k\sum\nolimits_{i \in I} {H'(i{d_M}||T_2||i){v_i}} }.
				\end{align}
				
				$\textbf{AggData}(params,\tilde{k}, \{On\text{-}Tag_{s}\}_{s\in S},\{M_s\}_{s\in S})$: Given a secret $\tilde{k}$, tags $\{On\text{-}Tag_{s}=(id_{M_s},\{t_{s,i}\}_{i=1}^\ell)\}_{s\in S}$, and data files $\{M_s\}_{s\in S}$, where $M_s=(m_{s,1},\ldots,m_{s,\ell})$, it computes     $m^{Agg}_i=\sum\nolimits_{s \in S} F_{\tilde{k}}(id_{M_s}||i) m_{s,i}$ and $t_i^{Agg} = \sum\nolimits_{s \in S} {{F_{\widetilde k}}(i{d_{{M_s}}}||i){t_{s,i}}}$, for  $i=1,\ldots, \ell$. It returns an aggregate data   		$M^{Agg}=(m^{Agg}_1,\ldots,m^{Agg}_\ell)$ and an aggregate tag  $On\text{-}Tag^{Agg}=(\{id_{M_s}\}_{s\in S},\{t_i^{Agg}\}_{i=1}^{\ell})$.
\vspace{0.15cm}
				\\			
				
				
				$\textbf{Offline\text{-}Challenge}(params,1^m,PK)$: Given a security parameter $m$ and a public key $PK=(\Lambda,\sigma)$, it selects $k_i\in \mathbb{Z}_q$ uniformly at random, for  $i=1,\ldots,m$. Then,   it computes $v^{(i)}=g^{k_i}$ and $w^{(i)}=\Lambda^{k_i}$, $i=1,\ldots,m$, and returns $PreChall=\{(k_i,v^{(i)},w^{(i)})\}_{i=1}^m$.
\vspace{0.15cm}
				\\
				
				$\textbf{ChallGen}_2(params,PreChall,N)$: On input  $PreChall=\{(k_i,v^{(i)},w^{(i)})\}_{i=1}^m$ and an integer $N$, it first selects $J_j\subset \{1,\ldots,m\}$ randomly, for $j=1,\ldots,N$. Then, for each $j=1,\ldots,N$, it computes ${k^{(j)}} = \sum\nolimits_{i \in {J_j}} {{k_i}}$,	$c_1^{(j)} = \prod\nolimits_{i \in {J_j}} {v^{(i)}}$, $c_2^{(j)} = \prod\nolimits_{i \in {J_j}} {w^{(i)}}$.
				Afterward, it considers a uniform subset $I\subseteq \{1,\ldots,\ell\}$ and   selects a uniform element $v_i \in \mathbb{Z}_p$, for each $i\in I$. This algorithm outputs a set of challenges $C_{Agg}=(\{C^{(j)}=(c_1^{(j)},c_2^{(j)})\}_{j=1}^{N},\{(i,v_i)\}_{i\in I})$ and the associated secretes $K_{Agg}=\{k^{(j)}\}_{j=1}^{N}$.
\vspace{0.15cm}
				\\
				
				
				
				
				
				$\textbf{CheckProof}_2(params, Off\text{-}Tag,C,k,\tilde{k},P^{Agg}, \{id_{M_s}\}_{s\in S})$: Given $Off\text{-}Tag=(T_1,T_2,y,\sigma',PK)$, $C=((c_1,c_2,\{(i,v_i)\}_{i\in I})$, secret keys $k$ and $\tilde{k}$, and $P^{Agg}=(P_1^{Agg},P_2^{Agg})$, this algorithm  returns $1$ if and only if the following equation holds:
				\begin{align}\label{4545}
					P_1^{Agg}P_2^{Agg} =\hat{e}{(H(ID),{T_1})^{k\sum\nolimits_{i \in I} {({v_i}\sum\nolimits_{s \in S} {{F_{\widetilde k}}(i{d_{{M_s}}}||i)H'(i{d_{{M_s}}}||{T_2}||i))} } }}.
				\end{align}
				
				$\textbf{CheckProof}_3(params, Off\text{-}Tag,C_{Agg},K_{Agg},\{P^{(j)}\}_{j=1}^N)$: On input $Off\text{-}Tag=(T_1,T_2,y,\sigma',PK)$, $C_{Agg}=(\{C^{(j)}=(c_1^{(j)},c_2^{(j)})\}_{j=1}^{N},\{(i,v_i)\}_{i\in I})$, $K_{Agg}=\{k^{(j)}\}_{j=1}^{N}$, and $\{P^{(j)}\}_{j=1}^N$, where $P^{(j)}=(P^{(j)}_1,P^{(j)}_2)$, this algorithm  returns $1$ if and only if the following equation holds:
				\begin{align}\label{45454}
					\prod\nolimits_{j = 1}^N {P_1^{(j)}P_2^{(j)}}  = \hat{e}(H(ID),T_1)^{\sum\nolimits_{j = 1}^N {{(k^{(j)}}} \sum\nolimits_{i \in I} {H'(i{d_M^{(j)}}||T_2||i)v_i)} }.
				\end{align}
				\vspace{0.15 cm}
				\\
					$\textbf{Loc\text{-}Trap}(params,\{id_{M_s}\}_{s\in S},C,k,SOff\text{-}Tag)$: Given  identifiers $\{id_{M_s}\}_{s\in S}$, a challenge $C=(c_1,c_2,\{(i,v_i)\}_{i\in I})$, a secret key $k$, and $SOff\text{-}Tag=(T_2,y)$, it computes $trap_s={k\sum\limits_{i \in I} {H'(i{d_{M_{s}}}||T_2||i){v_i}} }$, for each $s\in S$, and returns $Trap_{S}=\{trap_s\}_{s\in S}$.
				\\
				\vspace{0.15 cm}
				$\textbf{Find\text{-}Corrupted}(params, POff\text{-}Tag,\{On\text{-}Tag_{s}\}_{s\in S},\{M_{s}\}_{s\in S},C,Trap_S)$: At first, this algorithm  executes 	$(P_{s,1},P_{s,2})\leftarrow\textbf{GenProof}(params,C,On\text{-}Tag_s,PK,M_{s})$, for any $s\in S$, and returns $S'=\{s\in S:\,\,P_{s,1}P_{s,2}\ne \hat{e}(H(ID),T_1)^{trap_s}\}$.
			}
		\end{myframe}
	\end{textbox}
	\caption{Algorithms used in our proposed scheme.}\label{ajnaa}
\end{figure*}

%

\section{Security analysis}\label{kljhjk}
Let $\mathcal{A}$, $\mathcal{C}$, and $\Pi$ be a PPT adversary, a PPT challenger, and our proposed ${\text{O}^2\text{DI}}$, respectively.  Consider the following experiment.

\noindent	\textbf{The proof forging experiment } $\mathbf{Proof}$-$\mathbf{forge}_{\mathcal{A},\Pi}(n)$:

\noindent
	1) \textbf{Start}: $\mathcal{C}$ executes $(params,MSK)\leftarrow\textbf{Setup}(1^n)$  and gives the public parameters $params$ to $\mathcal{A}$. Also, it considers a list $L_{ID}$ which is initially empty.
	
\noindent	2) \textbf{Phase 1}: 
	$\mathcal{A}$  asks a number of  questions to the  oracle $\mathcal{O}_{\textbf{Extract}}(ID)$, and $\mathcal{C}$ answers them as follows: 
	\begin{itemize}
		\item $\mathcal{O}_{\textbf{Extract}}(ID)$: Given an identifier  $ID$,  $\mathcal{C}$ runs  \scalebox{0.98}{$(SK,PK)\leftarrow\textbf{Extract}(params,MSK,ID)$} and gives $(SK,PK)$ to $\mathcal{A}$. Also, $\mathcal{C}$ adds $ID$ to $L_{ID}$.
	\end{itemize}
	3) \textbf{Phase 2}: When \textbf{Phase 1} is over, $\mathcal{A}$ gives a challenge identifier $ID'$   to $\mathcal{C}$.   If  $ID' \in L_{ID}$, then $\mathcal{C}$  asks $\mathcal{A}$ to choose another challenge identifier. Otherwise, $\mathcal{C}$ runs $(SK',PK')\leftarrow\textbf{Extract}(params, MSK,ID')$ and $Off\text{-}Tag^*\leftarrow\textbf{Offline-Tag}(params,PK',\\SK')$. Then, it returns $PK'$ to $\mathcal{A}$. Afterward,   
	$\mathcal{A}$ makes a polynomial number of queries to the oracles $\mathcal{O}_{\textbf{Extract}}(ID)$ and $\mathcal{O}_{\textbf{Online-Tag}}(M)$, and $\mathcal{C}$ responds to them as follows:
	\begin{itemize}
		\item $\mathcal{O}_{\textbf{Extract}}(ID)$: For an identifier $ID$ chosen by  $\mathcal{A}$, if $ID\ne ID'$, then $\mathcal{C}$ executes the algorithm $(SK,PK)\leftarrow\textbf{Extract}(params,\\MSK,ID)$ and gives $(SK,PK)$ to $\mathcal{A}$. Otherwise, it does not respond to the query.
		
		{$\mathcal{O}_{\textbf{Online\text{-}Tag}}(M)$}: For a data file $M$,  $\mathcal{C}$  executes $On\text{-}Tag\leftarrow\textbf{Online\text{-}Tag}(params,id_M,Off\text{-}Tag^*,M,PK')$ and gives $On\text{-}Tag$ to $\mathcal{A}$. 
	\end{itemize}
	4) \textbf{Challenge}:  $\mathcal{C}$  selects  $M^*\in\mathbb{Z}_q^\ell$,   $i \in \{1,\ldots,\ell\}$, and $x \in \mathbb{Z}_q$   randomly. Assuming  $M^*=m^*_1||\ldots||m^*_\ell$, it considers  ${M^*}'=m^*_1||\ldots||m^*_{i-1}||x||m^*_{i+1}||\ldots||m^*_\ell$ and executes $On\text{-}Tag^*\leftarrow\textbf{Online-Tag}(params,id_M,Off\text{-}Tag^*,{M^*},SK')$,  $(C^*,k^*)\leftarrow\textbf{ChallGen}_1(params, PK')$. Finally, $\mathcal{C}$ returns $(On\text{-}Tag^*,{M^*}',C^*)$    to $\mathcal{A}$.  
	
\noindent	5) \textbf{Forge}: The adversary $\mathcal{A}$ returns a proof $P^*=(P^*_1,P^*_2)$ to the challenger.

We say that $\mathcal{A}$ succeeds, and we write \scalebox{0.85}{$\mathbf{Proof}$-$\mathbf{forge}_{\mathcal{A},\Pi}(n)=1$}, if we have $1=\textbf{CheckProof}_1(params,Off\text{-}Tag^*,C^*,k^*,P^*)$.

\begin{Remark}
	The experiment $\mathbf{Proof}$-$\mathbf{forge}_{\mathcal{A},\Pi}(n)$  described above has been designed specially for  non-aggregate data verification process. However, if we substitute  $\textbf{ChallGen}_2$ for $\textbf{ChallGen}_1$ in the \textbf{Challenge} stage of the experiment,  then we obtain a new experiment that accurately models  adversaries' objectives and abilities in the  aggregate verification process.  Let us denote the new experiment by $\mathbf{Proof}$-$\mathbf{forge}^{Agg}_{\mathcal{A},\Pi}(n)$.
\end{Remark}

Before defining the security of our ${\text{O}^2\text{DI}}$, we first show that  the  experiments $\mathbf{Proof}$-$\mathbf{forge}_{\mathcal{A},\Pi}(n)$ and  $\mathbf{Proof}$-$\mathbf{forge}^{Agg}_{\mathcal{A},\Pi}(n)$ are  equivalent. This  means that if the proof of possession provided through our ${\text{O}^2\text{DI}}$ scheme is existentially unforgeable in the non-aggregate verification process, then  it is  also  unforgeable  in the aggregate  verification process.   

\begin{Theorem}\label{l1}
	For any PPT adversary $\mathcal{A}$, there exists a PPT adversary $\mathcal{B}$ such that:
	{$$\Pr[\mathbf{Proof}\text{-}\mathbf{forge}^{Agg}_{\mathcal{B},\Pi}(n)=1]=\Pr[\mathbf{Proof}\text{-}\mathbf{forge}_{\mathcal{A},\Pi}(n)=1].$$}
\end{Theorem}

\begin{IEEEproof}
	Let $\mathcal{A}$ and $\mathcal{C}$ be an arbitrary  PPT adversary and a challenger in the experiment $\mathbf{Proof}\text{-}\mathbf{forge}_{\mathcal{A},\Pi}(n)$, respectively. Consider another  PPT adversary $\mathcal{B}$  in the experiment $\mathbf{Proof}\text{-}\mathbf{forge}_{\mathcal{B},\Pi}^{Agg}(n)$. Assume that  $\mathcal{B}$ is executed as a subroutine by $\mathcal{A}$ as follows: 
	
	\noindent 1)	
	\textbf{Start}: $\mathcal{C}$ runs \scalebox{0.97}{$(params,MSK)\leftarrow\textbf{Setup}(1^n)$}  and gives $params$ to $\mathcal{A}$.  Afterward, $\mathcal{A}$ returns $params$ to $\mathcal{B}$.
	\\	
	\noindent 2)
	 \textbf{Phase 1}: When the adversary $\mathcal{B}$  requests the secret key and public key associated with an identifier $ID$ from   $\mathcal{A}$, the adversary  $\mathcal{A}$ obtains the requested   keys $(SK,PK)$ through $\mathcal{O}_{\textbf{Extract}}(ID)$ and  gives them to  $\mathcal{B}$. 
	\\	
	\noindent 3)	
	\textbf{Phase 2}:  $\mathcal{B}$ chooses  an identifier  $ID'$, and $\mathcal{A}$ returns it to  $\mathcal{C}$.   Then, $\mathcal{C}$ executes  \scalebox{0.99}{$(SK',PK')\leftarrow\textbf{Extract}(params, MSK,ID')$} and $Off\text{-}Tag^*\leftarrow\textbf{Offline-Tag}(params,PK',SK')$ and gives  $PK'$ to $\mathcal{A}$. Afterward,  $\mathcal{A}$ returns $PK'$  to $\mathcal{B}$ and answers the queries made by $\mathcal{B}$ to the following  oracles: 
		
		\begin{itemize}
			\item $\mathcal{O}_{\textbf{Extract}}(ID)$: For an identifier $ID$ chosen by  $\mathcal{B}$,  if $ID\ne ID'$, then $\mathcal{A}$ calls $\mathcal{O}_{\textbf{Extract}}(ID)$ to obtain  $(SK,PK)$. It  gives $(SK,PK)$ to $\mathcal{B}$ and adds $ID$ to $L'_{ID}$.
			\item $\mathcal{O}_{\textbf{Online-Tag}}(M)$: For a data file $M$ selected by $\mathcal{B}$,  $\mathcal{A}$  calls    $\mathcal{O}_{\textbf{Online\text{-}Tag}}(M)$ and   gives $On\text{-}Tag$ to $\mathcal{B}$. 
		\end{itemize}
\noindent		4)
		  \textbf{Challenge}:   $\mathcal{C}$ 
		selects  $M^*\in\mathbb{Z}_q^\ell$,   $i \in \{1,\ldots,\ell\}$, and $x \in \mathbb{Z}_q$ and sets  ${M^*}'=M^*_1||\ldots||M^*_{i-1}||x||M^*_{i+1}||\ldots||M^*_m$, where {$M^*=M^*_1||\ldots||M^*_m$}. Then,  running $On\text{-}Tag^*\leftarrow\textbf{Online-Tag}(params,id_{M^*},Off\text{-}Tag^*,{M^*},SK')$, the challenger provides $\mathcal{A}$ with a tuple $(On\text{-}Tag^*,{M^*}',C^*)$, where $C^*=(c_1^*,c^*_2)$.   When  $\mathcal{A}$ receives the tuple, for an integer $N>1$, it first selects $k^{(1)},\ldots, k^{(N-1)}\in \mathbb{Z}_q$ uniformly at random and computes  $c_1^{(j)}=g^{-k^{(j)}}$ and $c_2^{(j)}=\Lambda^{-k^{(j)}}$, for {$j=1,\ldots, N-1$}, and $c_1^{(N)}=c_1^*\prod\nolimits_{j = 1}^{N-1} {c_1^{(j)}} $ and $c_2^{(N)}=c_2^*\prod\nolimits_{j = 1}^{N-1} {c_2^{(j)}} $. One can obviously  see that,   $\prod\nolimits_{j = 1}^N {c_1^{(j)}}=c_1^*$, $ \prod\nolimits_{j = 1}^N {c_2^{(j)}}=c_2^*$, and $k^* = \sum\nolimits_{j = 1}^N {{k^{(j)}}}$, where $k^*$ and $k^{(j)}$ are the secrets associated with $C^*$ and $(c_1^{(j)},c_{2}^{(j)})$, $j=1,\ldots,N$, respectively. 
\\		\noindent 5)  \textbf{Forge}: $\mathcal{B}$ returns $\{(P_{1}^{(j)},P_{2}^{(j)})\}_{j=1}^N$ to  $\mathcal{A}$, and  $\mathcal{A}$ returns $(P_1^*,P_2^*)$ to $\mathcal{C}$, where $P_1^* = \prod\nolimits_{j = 1}^N {P_1^{(1)}}$ and $P_2^* = \prod\nolimits_{j = 1}^N {P_2^{(1)}}$. 
	We see that if 	
	$\prod\nolimits_{j = 1}^N {P_1^{(j)}P_2^{(j)}}  = T_2^{\sum\nolimits_{j = 1}^N {{k^{(j)}}} \sum\nolimits_{i \in I} {H(i{d_f}||i)} }$,
	then we have 
	$P_1P_2=T_2^{k\sum\nolimits_{i \in I} {H(i{d_f}||i)v_i} }$.
	Thus,
	\begin{align}\nonumber
		\Pr[\mathbf{Proof}\text{-}\mathbf{forge}^{Agg}_{\mathcal{B},\Pi}(n)=1]=\Pr[\mathbf{Proof}\text{-}\mathbf{forge}_{\mathcal{A},\Pi}(n)=1].
	\end{align}
	
\end{IEEEproof}
Now, according to Theorem  \ref{l1}, we can define the security of our  ${\text{O}^2\text{DI}}$ as follows:
\begin{Definition}
	Our ${\text{O}^2\text{DI}}$ scheme $\Pi$ is said to be secure, if for all PPT adversaries $\mathcal{A}$ the value of $\Pr[\mathbf{Proof}$-$\mathbf{forge}_{\mathcal{A},\Pi}(n)=1]$ is a negligible function in $n$.
\end{Definition}

\begin{Theorem}
	If the BDH problem is hard relative to $\mathcal{G}$, then our ${\text{O}^2\text{DI}}$ scheme is secure in the random oracle model. 
\end{Theorem}	
\begin{IEEEproof}
	Let $\Pi$, $\mathcal{A}$, and $n$ denote our ${\text{O}^2\text{DI}}$, an arbitrary PPT adversary, and  a security parameter, respectively. We show that the value of $\Pr[\mathbf{Proof}$-$\mathbf{forge}_{\mathcal{A},\Pi}(n)=1]$ is a negligible function in $n$,  where	 $\mathbf{Proof}\text{-}\mathbf{forge}_{\mathcal{A},\Pi}(n)$ is the proof forging experiment  described at the beginning of this section.
	
	Consider another PPT adversary $\mathcal{B}$ aiming to solve the BDH problem.  Given a tuple $(n,q,G_1,G_2,\hat{e},g, g^\alpha,g^\beta,g^\gamma)$, $\mathcal{B}$ wants to compute $\hat{e}(g,g)^{\alpha\beta\gamma}$, where $(q,G_1,G_2,\hat{e})\leftarrow \mathcal{G}(1^n)$, $g \in G_1$ is a random generator, and  $\alpha,\beta,\gamma \in \mathbb{Z}_q$ are chosen randomly. To solve the BDH problem, $\mathcal{B}$ runs $\mathcal{A}$ as a subroutine as follows: 
	
\noindent	1) \textbf{Start}: $\mathcal{B}$  selects   $H:\{0,1\}^*\to G_1$, $H':\{0,1\}^*\to \mathbb{Z}_q$, $F:\mathbb{Z}_q\times\mathbb{Z}_q\to \mathbb{Z}_q$,  and a secure IBS scheme \scalebox{0.98}{$\Pi=(\textbf{Gen},\textbf{Extract},$}\\$\textbf{Sign},\textbf{Vrfy})$. It  chooses  $ID^*$ and executes $(MSK_s,params_s)\leftarrow \textbf{Gen}(1^n)$ and $sk^*\leftarrow\textbf{Extract}(Params_s,MSK_s,ID^*)$. Then, it picks a natural number $\ell$ and gives $params=(n,q,g,G_1,G_2,\hat{e},H,H',F,\Pi_s,ID^*,params_s,pk=g^\alpha,\ell)$   to $\mathcal{A}$.
		
\noindent		2) \textbf{Phase 1}: 
		$\mathcal{A}$  makes some queries to  oracles $O_{H}$, $O_{H'}$, and $\mathcal{O}_{\textbf{Extract}}(ID)$,  and  $\mathcal{B}$, considering two initially empty sets $S_H$  and $S_{H'}$,   responds to the queries as follows:  
		\begin{itemize}
			\item $\mathcal{O}_{H}(X)$: Given $X\in \{0,1\}^*$, $\mathcal{B}$ at first checks whether $S_H$ includes $(X,x)$ or not. If not, it first selects     $x\in \mathbb{Z}_q$ uniformly at random and adds $(X,x)$ to $S_H$. Finally, it returns $H(X)=(g^\alpha)^{x}$. 
			
			\item $\mathcal{O}_{H'}(X)$: For $X\in \{0,1\}^*$, if there is $(X,x)\in S_{H'}$, then   $\mathcal{B}$ returns $H'(X)=x$  to $\mathcal{A}$. Otherwise,  the adversary  $\mathcal{B}$ first chooses a uniform element $x\in \mathbb{Z}_q$  and sets  $H'(X)=x$. Then,  $\mathcal{B}$ substitutes   $S_{H'}\cup \{(X,x)\}$   for   $S_{H'}$.
			
			\item $\mathcal{O}_{\textbf{Extract}}(ID)$: For an identifier  $ID$ given  by $\mathcal{A}$, $\mathcal{B}$ first checks whether there exists an element  $(ID,x_{ID})$ in $S_H$ or not. If not, it first selects $x_{ID}\in \mathbb{Z}_q$ uniformly at random and adds $(ID,x_{ID})$ to $S_H$. Afterward,   $\mathcal{B}$ computes the values $H=(g^{\alpha})^{x_{ID}\eta}=H(ID)^{\eta}$, $\Lambda=g^{\lambda}$,  $sk\leftarrow\textbf{Extract}(Params_s,MSK_s,ID)$, and $\sigma\leftarrow\textbf{Sign}(params_s,sk^*,\Lambda)$, where $ \lambda\in \mathbb{Z}_q$ is selected uniformly at random. Finally, $SK=(\lambda,sk)$ and $PK=(\Lambda,\sigma)$  are given to   $\mathcal{A}$.
		\end{itemize}

		\noindent 3) \textbf{Phase 2}: $\mathcal{A}$ picks an  $ID'$  such that the secret key associated with $ID'$ has not been queried in \textbf{Phase 1}. $\mathcal{B}$   selects  uniform elements   $\lambda,y \in \mathbb{Z}_q$ and   sets $\Lambda=g^{\lambda}$, \scalebox{0.9}{$sk\leftarrow\textbf{Extract}(Params_s,MSK_s,ID')$},  {$\sigma\leftarrow\textbf{Sign}(params_s,sk^*,\Lambda)$}, $T_1=g^y$, and ${\sigma'}\leftarrow \textbf{Sign}(params_s,sk,T_1)$. Then,
		$\mathcal{B}$ returns $PK=(\Lambda,\sigma)$  to $\mathcal{A}$. Afterward, $\mathcal{A}$ makes a polynomial number of queries to the following oracles, and $\mathcal{B}$ answers  $\mathcal{A}$ as follows. 	Note that according to the notations used in Figure \ref{ajnaa},  we have  $y=\beta$, and thus $T_2=pk^y=g^{\alpha\beta}$.
		
		\begin{itemize}
			\item $\mathcal{O}_{H}(X)$ and $\mathcal{O}_{H'}(X)$: The adversary $\mathcal{B}$ responds to these oracles the same as in \textbf{Phase 1}.

			\item $\mathcal{O}_{\textbf{Extract}}(ID)$: If $ID= ID'$, then  $\mathcal{B}$ does not respond to the query. Otherwise, $\mathcal{B}$ generates the requested  keys the same as in \textbf{Phase 1}.
			
			\item $\mathcal{O}_{\textbf{Online-Tag}}(M)$: Given  $M=(m_1,\ldots,m_\ell)$, $\mathcal{B}$ first selects an identifier $id_M\in \{0,1\}^*$ and uniform elements $t_1,\ldots,t_\ell \in \mathbb{Z}_q$. Then, it returns    $On\text{-}Tag=(id_M,\{t_i\}_{i=1}^{\ell})$.
			Note\,\, that\,\, assuming {$H'(id_M||g^{\alpha\beta}||i)=y^{-1}(t_i-\lambda^*m_i)$},  {$i=1,\ldots,\ell$}, one can see that $On\text{-}Tag$ is a valid tag. 
			

		\end{itemize}
		
		\begin{Remark}\label{56563}
			It is vital to note that by the hardness assumption of the BDH problem, and thus the hardness of the DH problem, the adversary $\mathcal{A}$ does not query an element in the form of  $X=(id_M||g^{\alpha\beta}||i)$ to the oracle $\mathcal{O}_{H'}$, except with a negligible probability. Therefore, by the programmability feature of the random oracle model \cite{modernn}, we can assume that $H'(id_M^*||g^{\alpha\beta}||i)=y^{-1}(t_i-\lambda^*m_i)$.
		\end{Remark}
	\noindent	4) \textbf{Challenge}: At first,  $\mathcal{B}$ chooses  $i^* \in \{1,\ldots,\ell\}$,  $x \in \mathbb{Z}_q$, and $I\subset \{1,\ldots,\ell\}$ uniformly at random. Then, it considers an unknown  data file $M^*=(m^*_1,\ldots,m^*_{i^*-1},\beta,m^*_{i^*+1},m^*_{\ell})$, where $m^*_j\in \mathbb{Z}_q$ is selected uniformly at random for $j\in \{1,\ldots,\ell\}\backslash \{i\}$. It also sets ${M^*}'=(m^*_1,\ldots,m^*_{i^*-1},x,m^*_{i^*+1},\ldots,m^*_{\ell})$, $c_1=g^{\gamma}$, $c_2=(g^{\gamma})^{\lambda}$, and $C^*=(c_1,c_2,\{v_i\}_{i\in I})$ and returns $(On\text{-}Tag^*,{M^*}',C^*)$ to $\mathcal{A}$, where $On\text{-}Tag^*=(id_{M^*},\{t_i^*\}_{i=1}^\ell)$ is the online tag associated with $M^*$.
		Note that according to the notations employed in Figure \ref{ajnaa}, in this case, we have $k=\gamma$.
		
	\noindent	5) \textbf{Forge}:   $\mathcal{A}$ returns a proof $P^*=(P^*_1,P^*_2)$.
	Upon receiving $P^*=(P^*_1,P^*_2)$, the adversary $\mathcal{B}$ first computes
	\\
	\scalebox{1}{$E=\hat e{({g^\alpha },{g^\gamma })^{ - ({v_{{i^*}}}{x_{ID'}}{t_{{i^*}}} + {v_{{i^*}}}{x_{ID'}}\sum\nolimits_{i \in I\backslash \{ {i^*}\} } {H'(i{d_M}||{T_2}||i){v_i})} }}$}
	and outputs $ {(P_1^*P_2^*E)^{ - {{({v_{{i^*}}}{x_{ID'}}\lambda )}^{ - 1}}}}$.

	Note that as we mentioned in Remark \ref{56563}, we have assumed that $H'(id_M^*||g^{\alpha\beta}||i^*)=y^{-1}(t_i-\lambda^*\beta)$. Therefore, if $\mathcal{A}$  succeeds in forging an integrity proof, then we have
	\begin{align}\nonumber
	P_1^*P_2^* &= \hat e{(H(ID'),{T_1})^{k\sum\nolimits_{i \in I} {H'(i{d_M}||{T_2}||i){v_i}} }}\\\nonumber
		&= \hat e{({g^{\alpha {x_{ID'}}}},{g^y})^{\gamma \sum\nolimits_{i \in I} {H'(i{d_M}||{T_2}||i){v_i}} }}\\\nonumber
		&= \hat e{({g^\alpha },{g^\gamma })^{{x_{ID'}}y\sum\nolimits_{i \in I} {H'(i{d_M}||{T_2}||i){v_i}} }}\\\nonumber
		&= \hat e{({g^\alpha },{g^\gamma })^{{x_{ID'}}y(H'(i{d_M}||{T_2}||{i^*}){v_{{i^*}}} + \sum\nolimits_{i \in I\backslash \{ {i^*}\} } {H'(i{d_M}||{T_2}||i){v_i}} )}}\\\nonumber
		&= \hat e{({g^\alpha },{g^\gamma })^{{x_{ID'}}y({v_{{i^*}}}{y^{ - 1}}({t_{{i^*}}} - \lambda \beta ) + \sum\nolimits_{i \in I\backslash \{ {i^*}\} } {H'(i{d_M}||{T_2}||i){v_i}} )}}\\\nonumber
		&= \hat e{({g^\alpha },{g^\gamma })^{ - {v_{{i^*}}}{x_{ID'}}\lambda \beta }}\hat E^{-1}\\\nonumber
		&= {(\hat e{(g,g)^{\alpha \beta \gamma }})^{ - {v_{{i^*}}}{x_{ID'}}\lambda }} E^{-1}.
	\end{align}
	\normalsize Therefore,  $ {(P_1^*P_2^*E)^{ - {{({v_{{i^*}}}{x_{ID'}}\lambda )}^{ - 1}}}}=\hat{e}(g,g)^{\alpha\beta\gamma}$,
	and thus
	\begin{align}\nonumber
		&\Pr[\mathbf{Proof}\text{-}\mathbf{forge}_{\mathcal{A},\Pi}(n)=1]\le \Pr [B(n,q,{G_1},{G_2},\hat e,g,{g^\alpha },{g^\beta },{g^\gamma }) = \hat e{(g,g)^{\alpha \beta \gamma }}].
	\end{align}
	On the other hand, by the hardness assumption of the BDH problem, we have $\Pr [\mathcal{B}(n,q,{G_1},{G_2},\hat e,g,{g^\alpha },{g^\beta },{g^\gamma }) = \hat e{(g,g)^{\alpha \beta \gamma }}]$ is  a negligible function in $n$. Thus,  $\Pr[\mathbf{Proof}\text{-}\mathbf{forge}_{\mathcal{A},\Pi}(n)=1]$ is also a negligible function. This completes the proof.

	%
\end{IEEEproof}

\begin{table*}[h]
	\caption{Comparison of computational, storage and communication overhead incurred by different schemes.}\label{taa}
	\centering
	\scalebox{0.85}{	\begin{tabular}{l|c|c|c|c|c|c|c|} 
			\hline  \myalign{|l}{Overhead} &\myalign{|l}{${\text{O}^2\text{DI}}$} &	\myalign{|l}{Shang \textit{et al.} \cite{dynamic} } & 	\myalign{|l}{Yu \textit{et al.}  \cite{Yuu} } &  \myalign{|l}{Li \textit{et al.}  \cite{certificate-based} } & \myalign{|l|}{Shu \textit{et al.}  \cite{Blockchainnn} }& \myalign{|l|}{Li \textit{et al.}  \cite{LI} } \\
			\hline
			\hline
			\myalign{|l}{{\multirow{2}{*}{$\mathcal{O}_1$}}}& 	\myalign{|l}{\multirow{2}{*}{$\ell(2 T_{M_z}+T_H)$}} & \myalign{|l}{$\ell(2T{e_1}+T_H)+$}&\myalign{|l}{$\ell(2T{e_1}+T_H)+$} & \myalign{|l}{{$\ell(T_H+2T_{e_1})+$}} & \myalign{|l|}{\multirow{2}{*}{$\ell(2T_H+3T_{e_1})+T_{e_1}$}}& {\multirow{2}{*}{$--$}} \\
			\myalign{|l}{}& 	\myalign{|l}{} 	& \myalign{|l}{$T_{e_1}+T_{Sign}$}&\myalign{|l}{$T_{e_1}+T_{Sign}$} &\myalign{|l|}{$3T_{e_1}+T_H$} &  \myalign{|l}{}& \myalign{|l|}{}\\\hline
			\myalign{|l}{\multirow{2}{*}{$\mathcal{O}_2$}}& \myalign{|l}{\multirow{2}{*}{$Nm(T_{A}+2T_{M_{g}})$}} &  \myalign{|l}{$N|S|(T_{e_1}+T_{e_2}+T_p)+$ } &\myalign{|l}{$N|S|(2T_{e_1}+2T_{e_2}+T_p)+$ }&  \myalign{|l}{\multirow{2}{*}{$N|S||I|T_{z\leftarrow \mathbb{Z}_q}$}}  & \myalign{|l|}{\multirow{2}{*}{$N|S||I|T_{z\leftarrow \mathbb{Z}_q}$}}& \myalign{|l|}{\multirow{2}{*}{$2N|S|T_{z\leftarrow\mathbb{Z}}$}}\\ 
			\myalign{|l}{}& \myalign{|l}{}	&  \myalign{|l}{$N|S|(|I|T_{z\leftarrow\mathbb{Z}_q}+T_H)$ } &\myalign{|l|}{$N|S|(|I|T_{z\leftarrow\mathbb{Z}_q}+T_H)$ }& \myalign{|l|}{}& \myalign{|l|}{}& \myalign{|l|}{{}}\\ \hline
			\myalign{|l}{\multirow{2}{*}{$\mathcal{O}_3$}}&\myalign{|l}{\multirow{2}{*}{$4T_p+2T_{e_1}+2|S|T_F$}} &  \myalign{|l}{\multirow{2}{*}{$|S|(|I| T_{e_1}+T_{e_2}+T_p)$}} &   \myalign{|l}{$|S|(|I| T_{e_1}+T_{e_2}+2T_p)+$}& \myalign{|l}{\multirow{2}{*}{$|S||I|T_{e_1}$}} & \myalign{|l|}{\multirow{2}{*}{$|S|(T_H+(2|I|+1)T_{e_1})$}} &\myalign{|l|}{$N|S|T_{z\leftarrow \mathbb{Z}_q}+$ }\\
			\myalign{|l}{}&\myalign{|l}{}	&  \myalign{|l}{} &   \myalign{|l}{$|S|(T_H+T_{Sign})$}& \myalign{|l|}{}& \myalign{|l|}{}&  \myalign{|l|}{{$N|S|(T_H+T_{e_1})$}}\\\hline
			\myalign{|l}{\multirow{2}{*}{$\mathcal{O}_4$}}&		\myalign{|l}{$(|I|||S|+|I|)T_{M_z}+$}& \myalign{|l}{{$|I| |S|(T_{e_1}+T_{H}+T_p)+$}}&\myalign{|l}{{$|I| |S|(T_{e_1}+T_{H}+T_p)+$} } &  \myalign{|l|}{$|S|(4T_p+3T_{H}+T_{e_1})+$}   & \myalign{|l|}{$3|S|(T_p+T_{e_1})+$}&\myalign{|l|}{\multirow{2}{*}{$|S|(T_H+2T_{e_1})$} }\\
			\myalign{|l}{}&\myalign{|l}{$|I||S|(T_H+T_F)+T_p$}		& \myalign{|l}{$|S|T_{e_1}$} &\myalign{|l}{$|S|(T_{e_1}+ T_{Vrfy})$}&\myalign{|l}{$|S||I|(T_{e_1}+T_{H})$}& \myalign{|l|}{$2|S||I|(T_H+T_{e_1})$}&\myalign{|l|}{}\\\hline
			\myalign{|l}{\multirow{2}{*}{$\mathcal{O}_5$}}& \myalign{|l}{\multirow{2}{*}{$\ell l_{\mathbb{Z}_q}$}} & \myalign{|l}{\multirow{2}{*}{$\ell l_{G_1}+ l_{Sign}+2l_{H}$}}  & \myalign{|l}{\multirow{2}{*}{$(\ell+1)l_{G_1}+l_{Sign}$}} & \myalign{|l}{\multirow{2}{*}{$(\ell+3)l_{G_1}$}} & \myalign{|l|}{\multirow{2}{*}{$2\ell l_{G_1}$}}  & {\multirow{2}{*}{$--$} }  \\
			\myalign{|l}{}& \myalign{|l}{} & \myalign{|l}{}  & \myalign{|l}{} & \myalign{|l}{} &\myalign{|l|}{}  & \myalign{|l|}{}  \\ \hline                                      
			\myalign{|l}{\multirow{2}{*}{$\mathcal{O}_6$}}& \myalign{|l}{\multirow{2}{*}{$2l_{G_1}+|I|l_{\mathbb{Z}_q}$}} & \myalign{|l}{$|S|(l_{G_1}+l_{G_2})+$}  & \myalign{|l}{\multirow{2}{*}{$4|S|l_{G_1}+|S|(|I|+1)l_{\mathbb{Z}_q}$}}  & \myalign{|l}{\multirow{2}{*}{$|I||S|l_{\mathbb{Z}_q}$}}  & \myalign{|l|}{\multirow{2}{*}{$|I||S|l_{\mathbb{Z}_q}$}} & \myalign{|l|}{\multirow{2}{*}{$|S|(l_{G_1}+l_{\mathbb{Z}_q})$} }  \\
			\myalign{|l}{}& \myalign{|l}{} & \myalign{|l}{$|I||S| l_{\mathbb{Z}_q}$}  & \myalign{|l}{} &\myalign{|l}{}  &  \myalign{|l|}{}   &\myalign{|l|}{}\\\hline
			\myalign{|l}{\multirow{2}{*}{$\mathcal{O}_7$}}& \myalign{|l}{\multirow{2}{*}{$2l_{G_2}$}} & \myalign{|l}{\multirow{2}{*}{$|S|(l_{G_1}+2l_H)$}}  & \myalign{|l}{\multirow{2}{*}{$|S|(l_{G_1}+l_{Sign}+l_H)$}} &\myalign{|l}{\multirow{2}{*}{$|S|(l_{\mathbb{Z}_q}+l_{G_1})$}} & \myalign{|l|}{\multirow{2}{*}{$|S|(l_{\mathbb{Z}_q}+3l_{G_1})$}}  & \myalign{|l|}{\multirow{2}{*}{$3|S|l_{G_1}$} }  \\
			\myalign{|l}{}&  \myalign{|l}{}& \myalign{|l}{}  & \myalign{|l}{}  &\myalign{|l}{} &  \myalign{|l|}{} & \myalign{|l|}{} \\ \hline
			\myalign{|l}{\multirow{2}{*}{$\mathcal{O}_8$}}& \myalign{|l}{\multirow{2}{*}{$\ell l_{\mathbb{Z}_q}$}} & \myalign{|l}{\multirow{2}{*}{$\ell l_{G_1}+ l_{Sign}+2l_{H}$}}  & \myalign{|l}{\multirow{2}{*}{$(\ell+1)l_{G_1}+l_{Sign}$}} & \myalign{|l}{\multirow{2}{*}{$(\ell+3)l_{G_1}$}} & \myalign{|l|}{\multirow{2}{*}{$2\ell l_{G_1}$}}  & {\multirow{2}{*}{$|M|$} }  \\
			\myalign{|l}{}& \myalign{|l}{} & \myalign{|l}{}  & \myalign{|l}{} & \myalign{|l}{} &\myalign{|l|}{}  & \myalign{|l|}{}  \\ \hline  
			\hline
			\multicolumn{7}{c}{\textbf{Notes.} \scalebox{1}{$\mathcal{O}_1$: Tag generation time; $\mathcal{O}_2$: Challenge generation time; $\mathcal{O}_3$: Response generation time; $\mathcal{O}_4$: Integrity auditing time; $\mathcal{O}_5$: Tag size; $\mathcal{O}_6$: Challenge size;  $\mathcal{O}_7$: Response size; }}\\
			\multicolumn{7}{c}{\scalebox{1}{$\mathcal{O}_8$: Storage cost incurred by the AV for each original data file.\,\,\,\,\,\,\,\,\,\,\,\,\,\,\,\,\,\,\,\,\,\,\,\,\,\,\,\,\,\,\,\,\,\,\,\,\,\,\,\,\,\,\,\,\,\,\,\,\,\,\,\,\,\,\,\,\,\,\,\,\,\,\,\,\,\,\,\,\,\,\,\,\,\,\,\,\,\,\,\,\,\,\,\,\,\,\,\,\,\,\,\,\,\,\,\,\,\,\,\,\,\,\,\,\,\,\,\,\,\,\,\,\,\,\,\,\,\,\,\,\,\,\,\,\,\,\,\,\,\,\,\,\,\,\,\,\,\,\,\,\,\,\,\,\,\,\,\,\,\,\,\,\,\,\,\,\,\,\,\,\,\,\,\,\,\,\,\,\,\,\,\,\,\,\,\,\,\,\,\,\,\,\,\,\,\,\,\,\,\,\,\,\,\,\,\,\,\,\,\,\,\,\,\,\,\,\,\,\,\,\,\,\,\,\,\,\,\,\,\,\,\,\,\,\,\,\,\,\,\,\,\,\,\,\,\,\,\,\,\,\,\,\,\,\,\,\,\,\, }}\\
	\end{tabular}}
	\label{tab:f1}
\end{table*}

\setcounter{figure}{0} 
\begin{figure*}[t]
	\multifig
	\centering
	\begin{tabular}{cccc}
		\figtarget & \figtarget & \figtarget& \figtarget \\
		\includegraphics[width=1.65in]{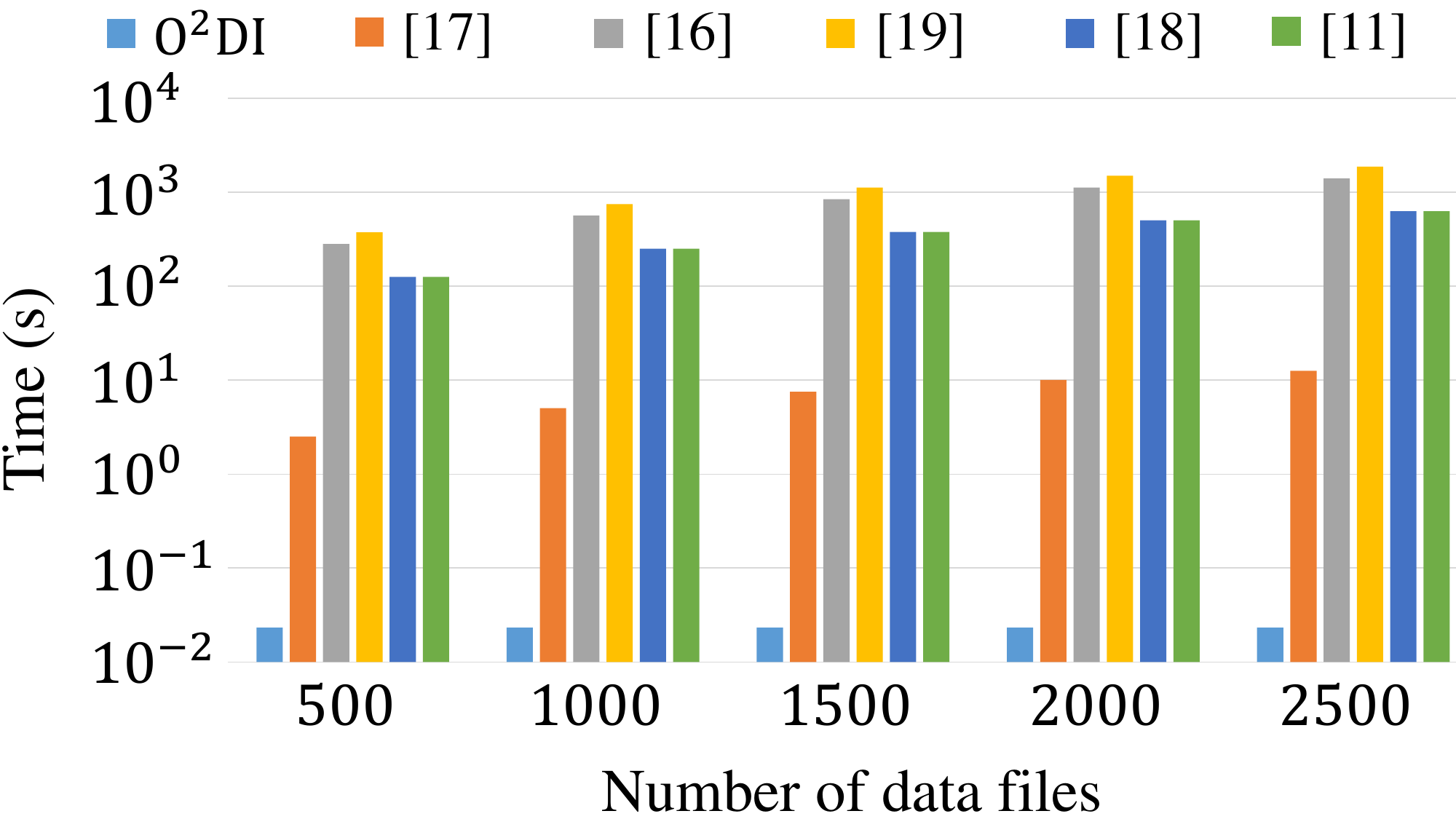} \captionsetup{labelformat=empty} &
		\includegraphics[width=1.65in]{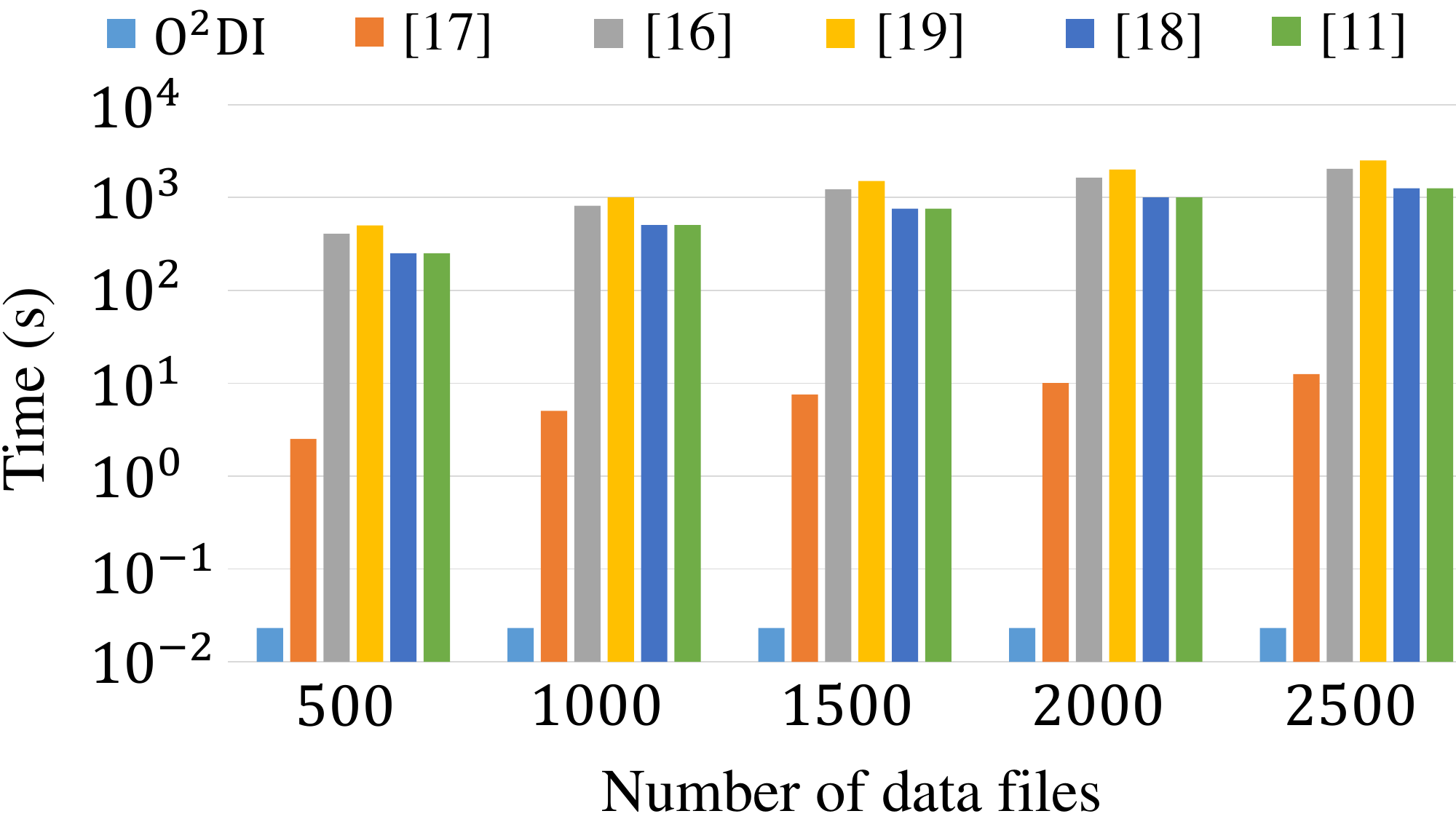} \captionsetup{labelformat=empty}&
		\includegraphics[width=1.65in]{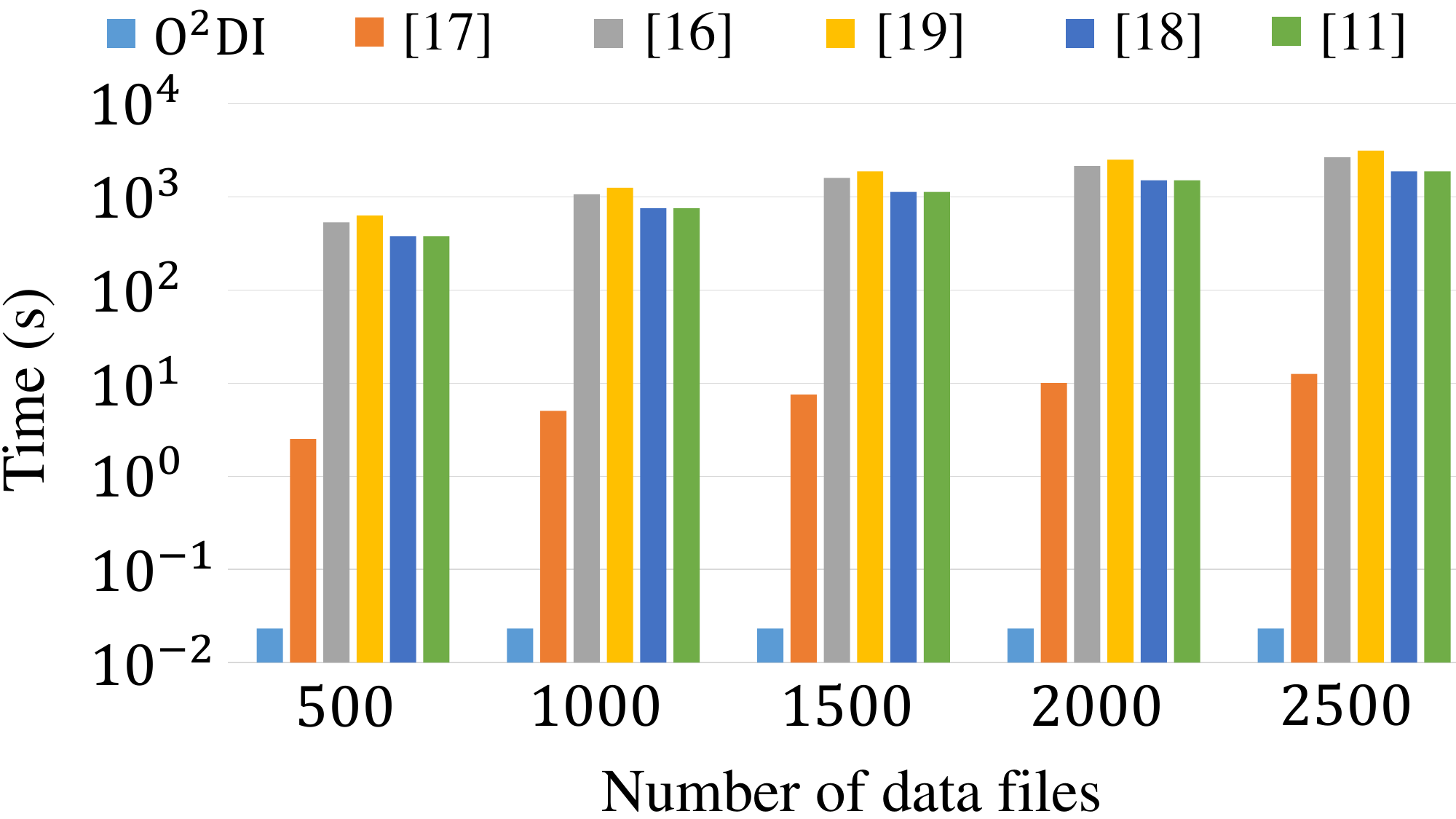} \captionsetup{labelformat=empty}&
		\includegraphics[width=1.65in]{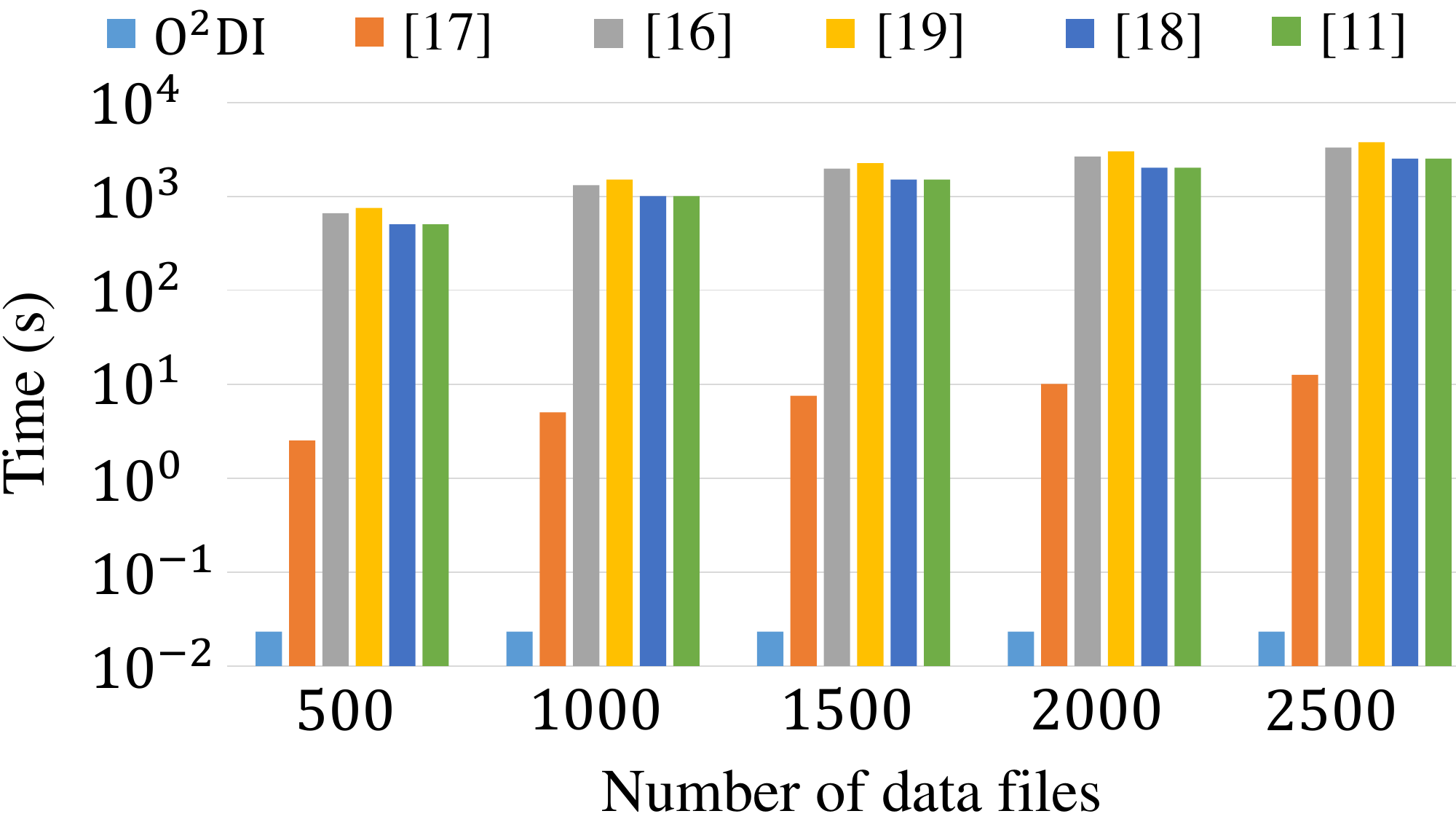} \\
		\scalebox{0.8}{(a) $I=100$} & \scalebox{0.8}{(b) $I=200$} &  \scalebox{0.8}{(c) $I=300$} & \scalebox{0.8}{(d) $I=400$}\\
		
	\end{tabular}
	\caption{The computational overhead on the app vendor side for generating   challenges.  }
	\label{ChallengeGenTimee}
\end{figure*}

\setcounter{figure}{1} 
\begin{figure*}[t]
	\multifig
	\centering
	\begin{tabular}{cccc}
		\figtarget & \figtarget & \figtarget& \figtarget \\
		\includegraphics[width=1.65in]{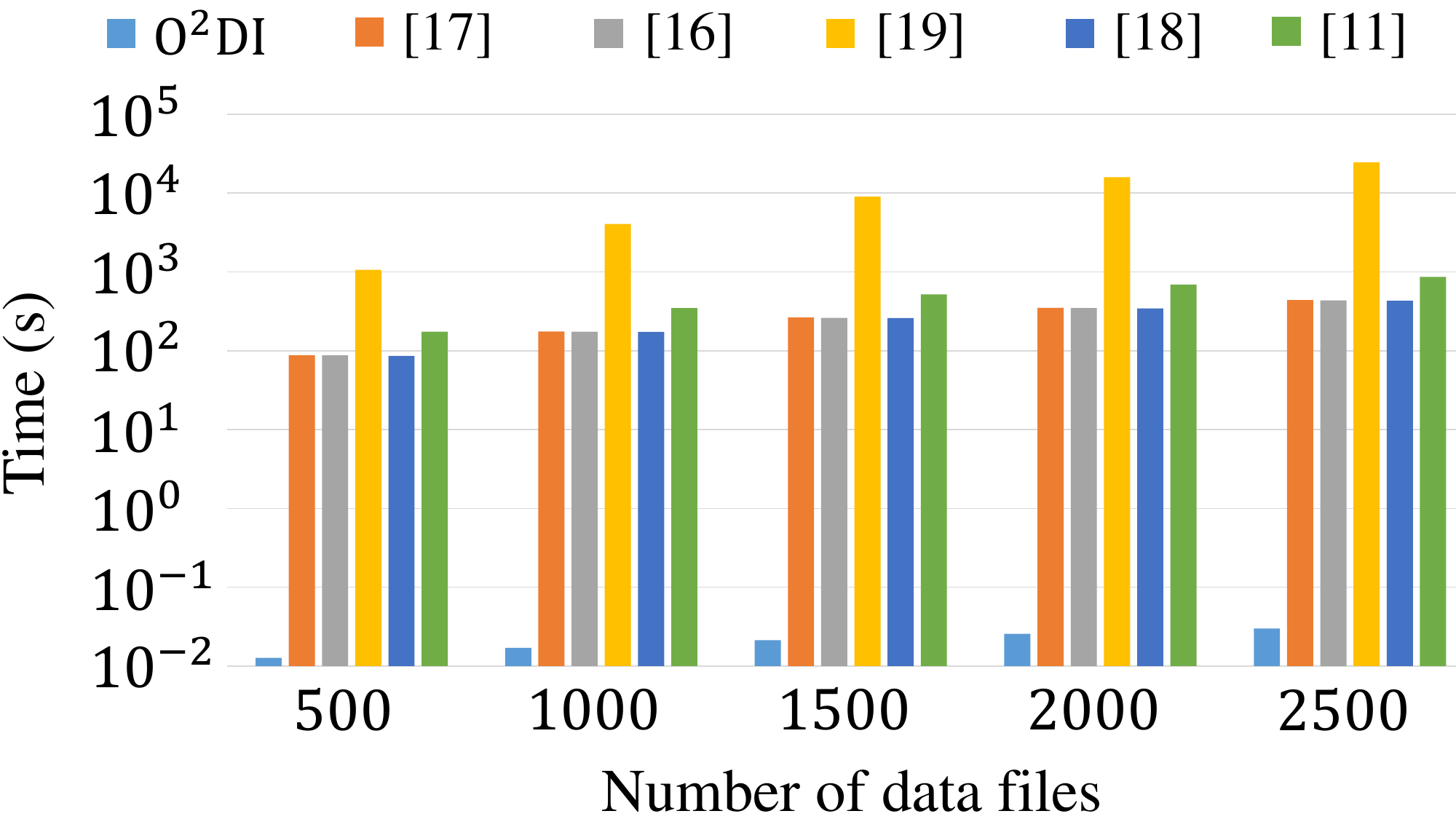} \captionsetup{labelformat=empty} &
		\includegraphics[width=1.65in]{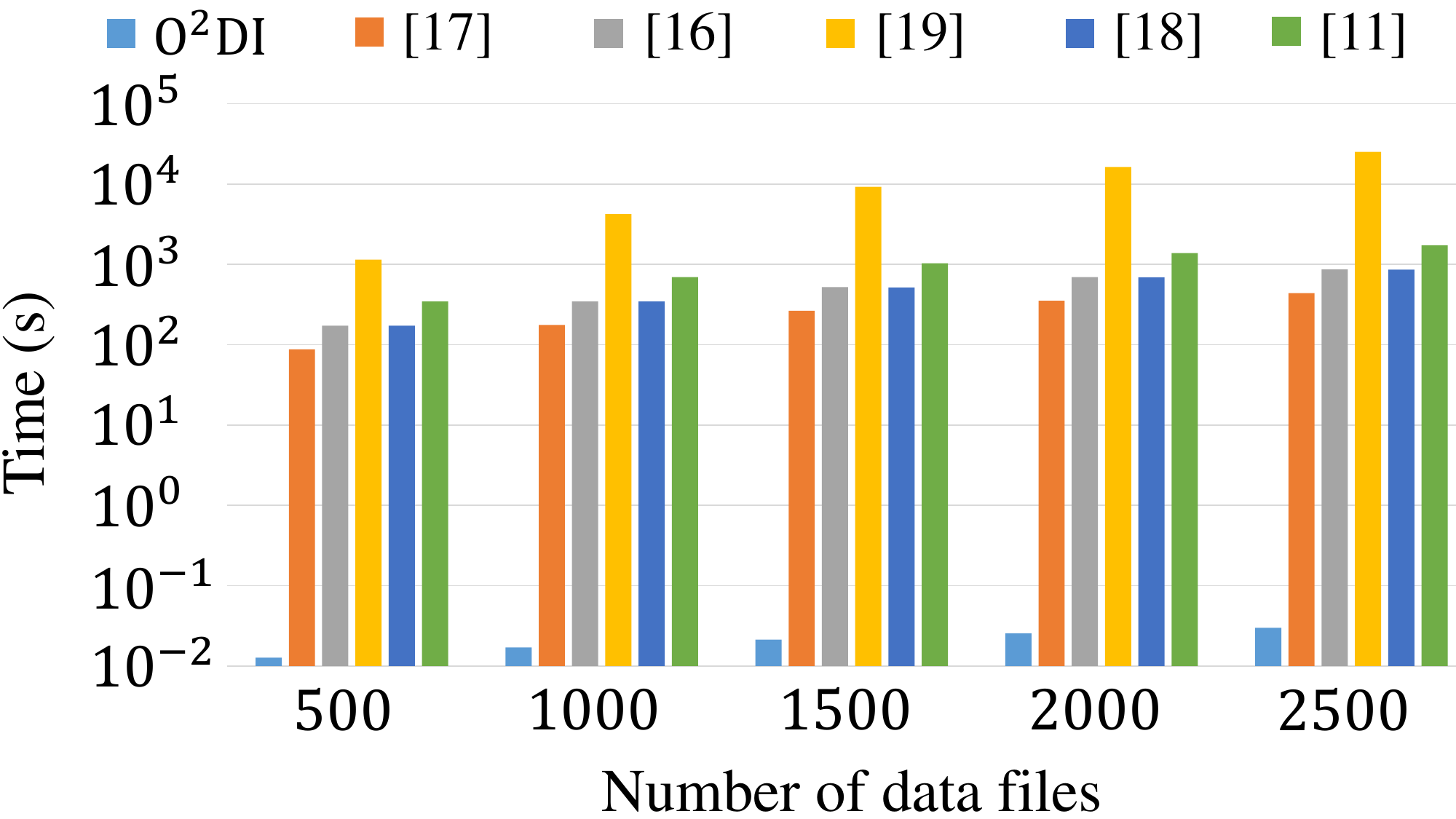} \captionsetup{labelformat=empty}&
		\includegraphics[width=1.65in]{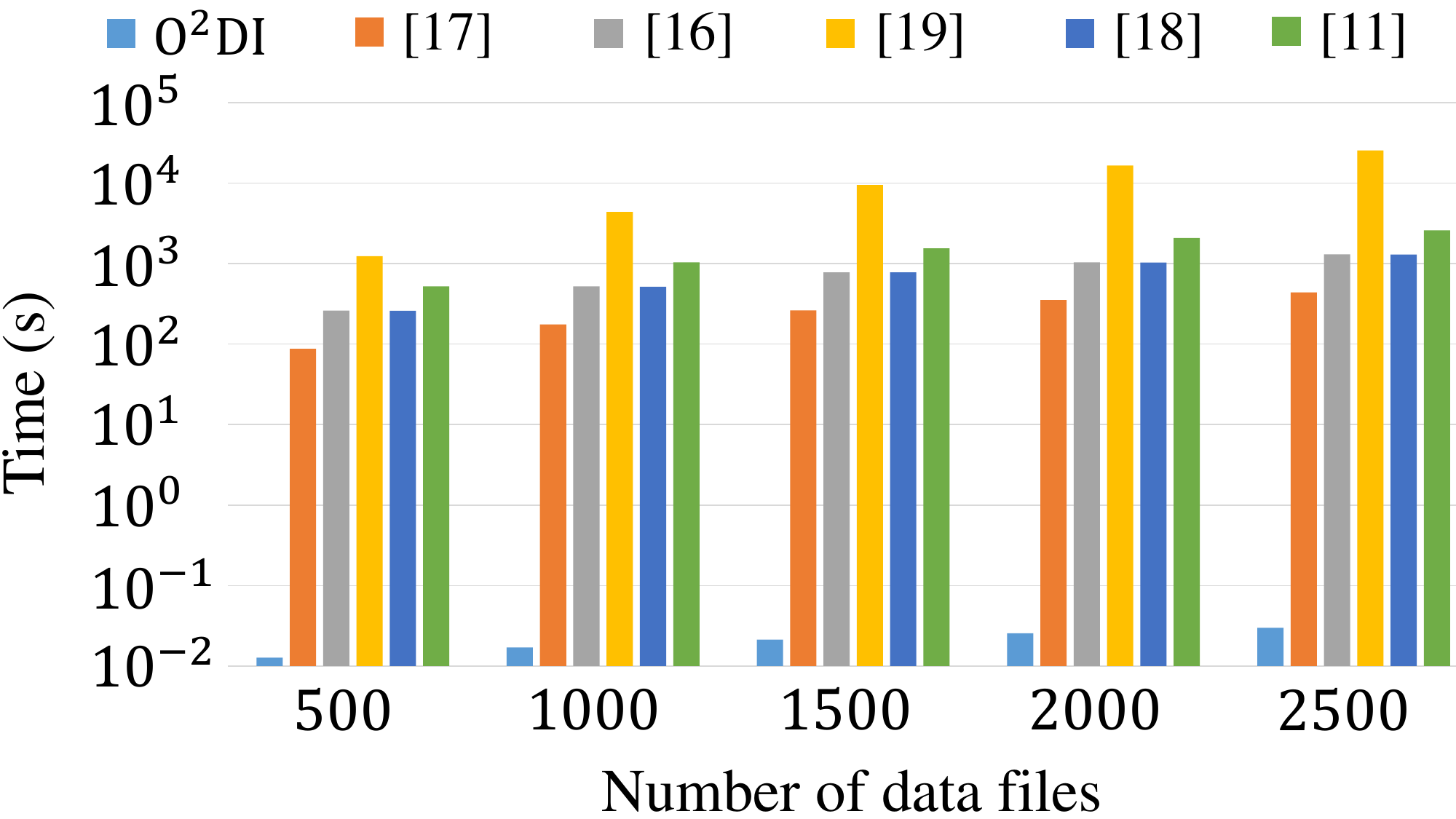} \captionsetup{labelformat=empty}&
		\includegraphics[width=1.65in]{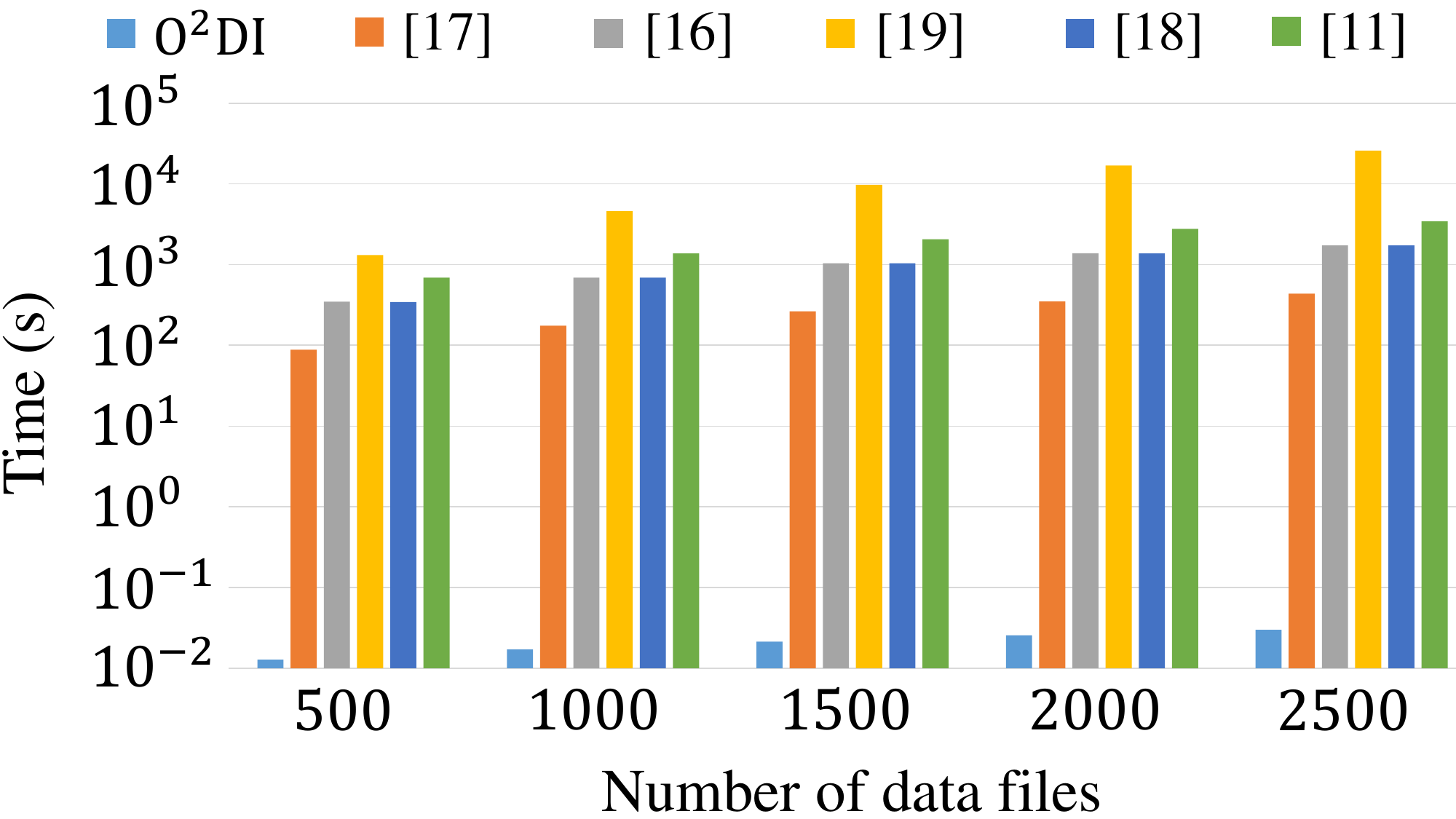} \\
		\scalebox{0.8}{(a) $I=100$} & \scalebox{0.8}{(b) $I=200$} &  \scalebox{0.8}{(c) $I=300$} & \scalebox{0.8}{(d) $I=400$}\\
		
	\end{tabular}
	\caption{The computational overhead on ESs for generating  an integrity proof.}
	\label{ProofGenTime}
\end{figure*}

\setcounter{figure}{2} 
\begin{figure*}[t]
	\multifig
	\centering
	\begin{tabular}{cccc}
		\figtarget & \figtarget & \figtarget& \figtarget \\
		\includegraphics[width=1.65in]{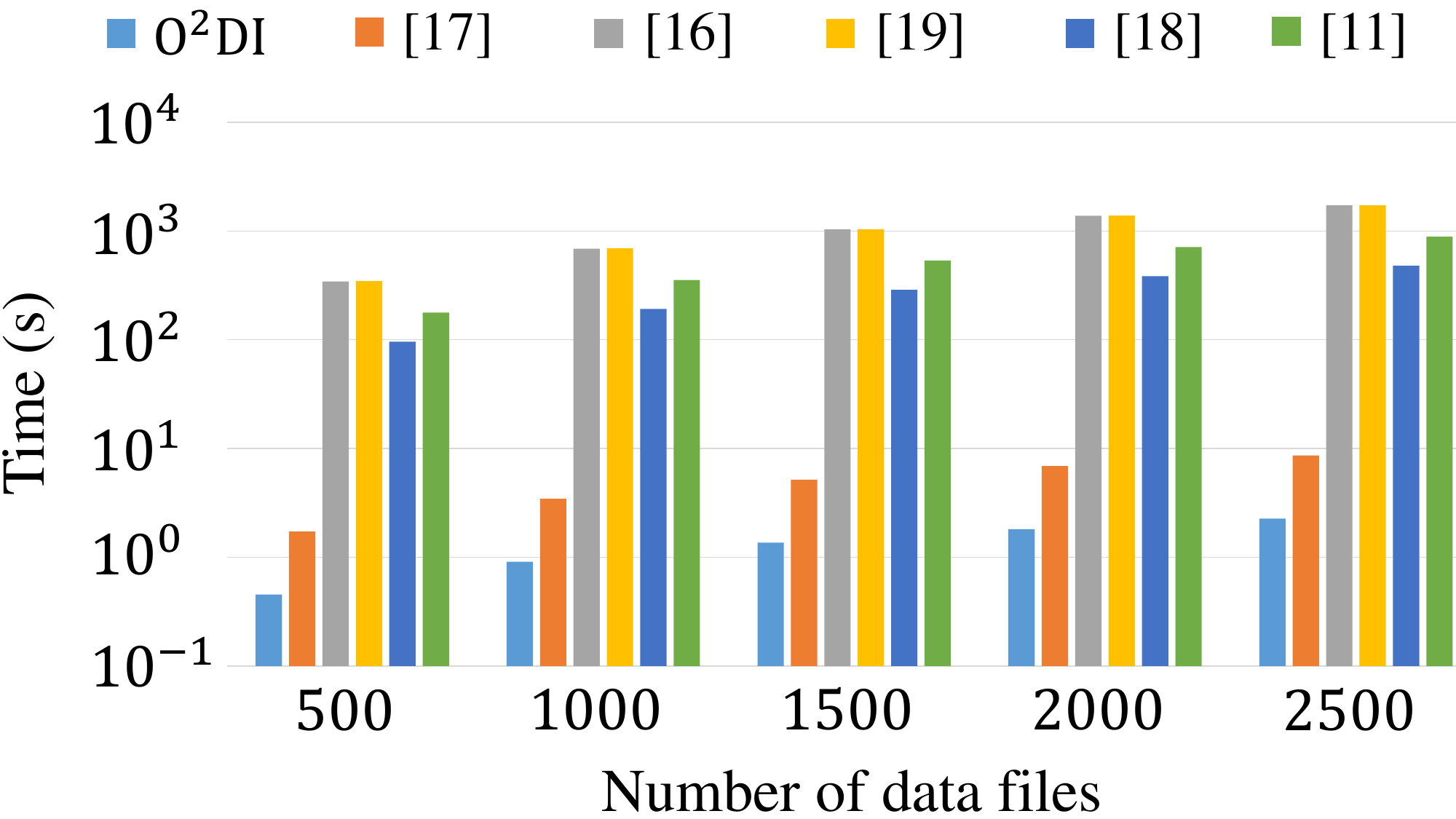} \captionsetup{labelformat=empty} &
		\includegraphics[width=1.65in]{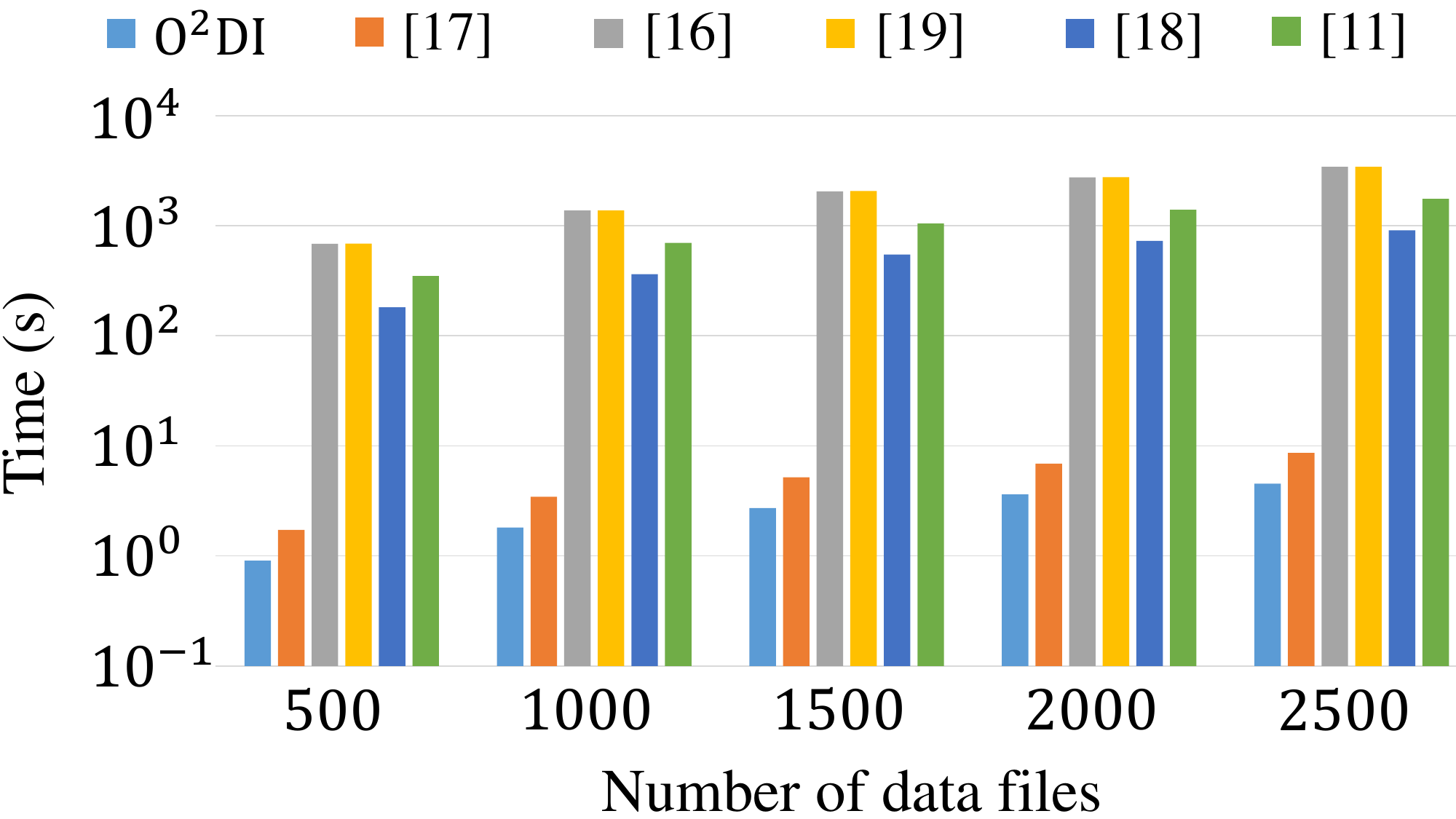} \captionsetup{labelformat=empty}&
		\includegraphics[width=1.65in]{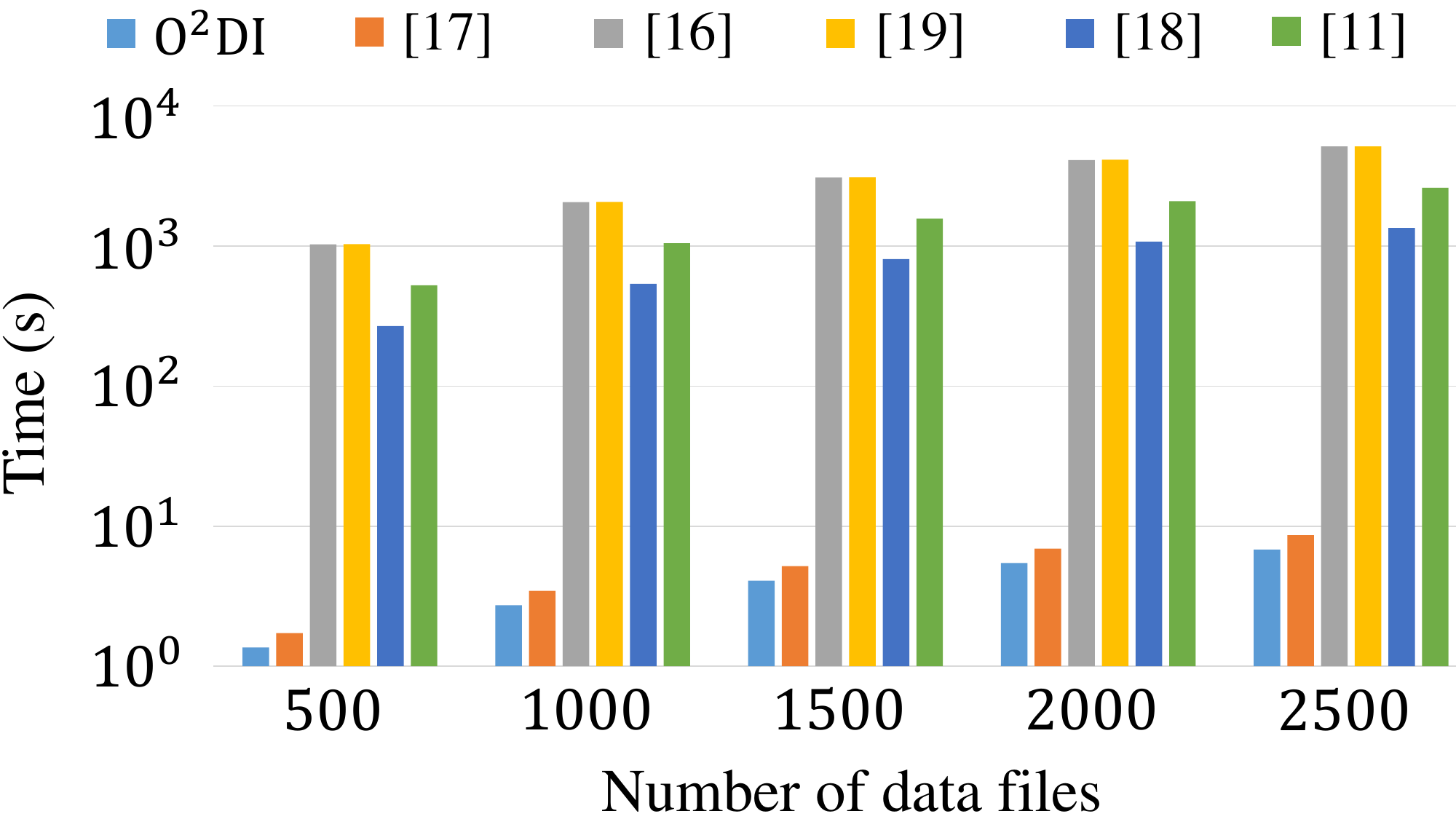} \captionsetup{labelformat=empty}&
		\includegraphics[width=1.65in]{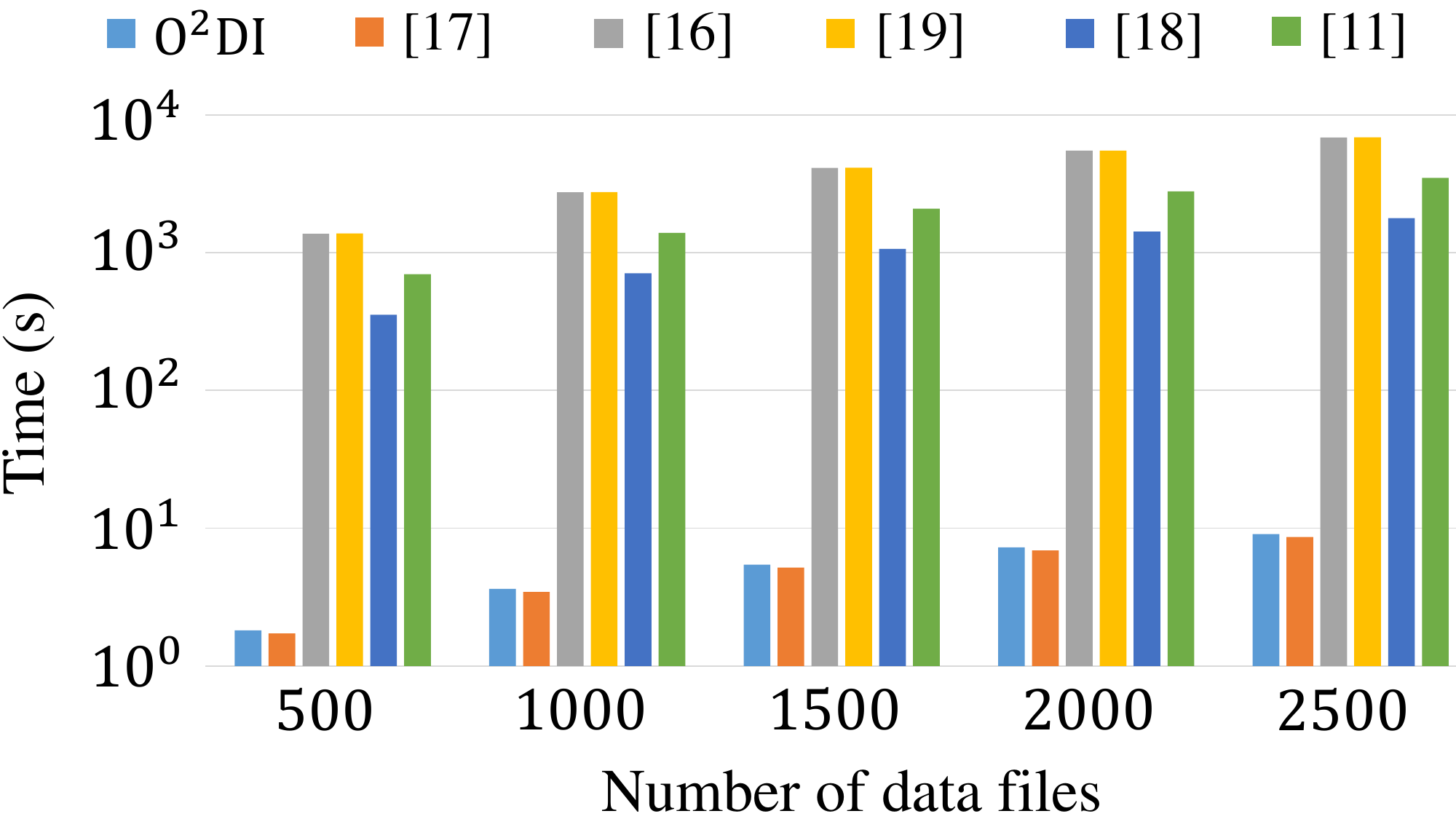} \\
		\scalebox{0.8}{(a) $I=100$} & \scalebox{0.8}{(b) $I=200$} &  \scalebox{0.8}{(c) $I=300$} & \scalebox{0.8}{(d) $I=400$}\\
		
	\end{tabular}
	\caption{The computational overhead on the app vendor for checking integrity proofs. }
	\label{IntCheckTime}
\end{figure*}

\setcounter{figure}{3} 
\begin{figure*}[t]
	\multifig
	\centering
	\begin{tabular}{cccc}
		\figtarget & \figtarget & \figtarget& \figtarget \\
		\includegraphics[width=1.65in]{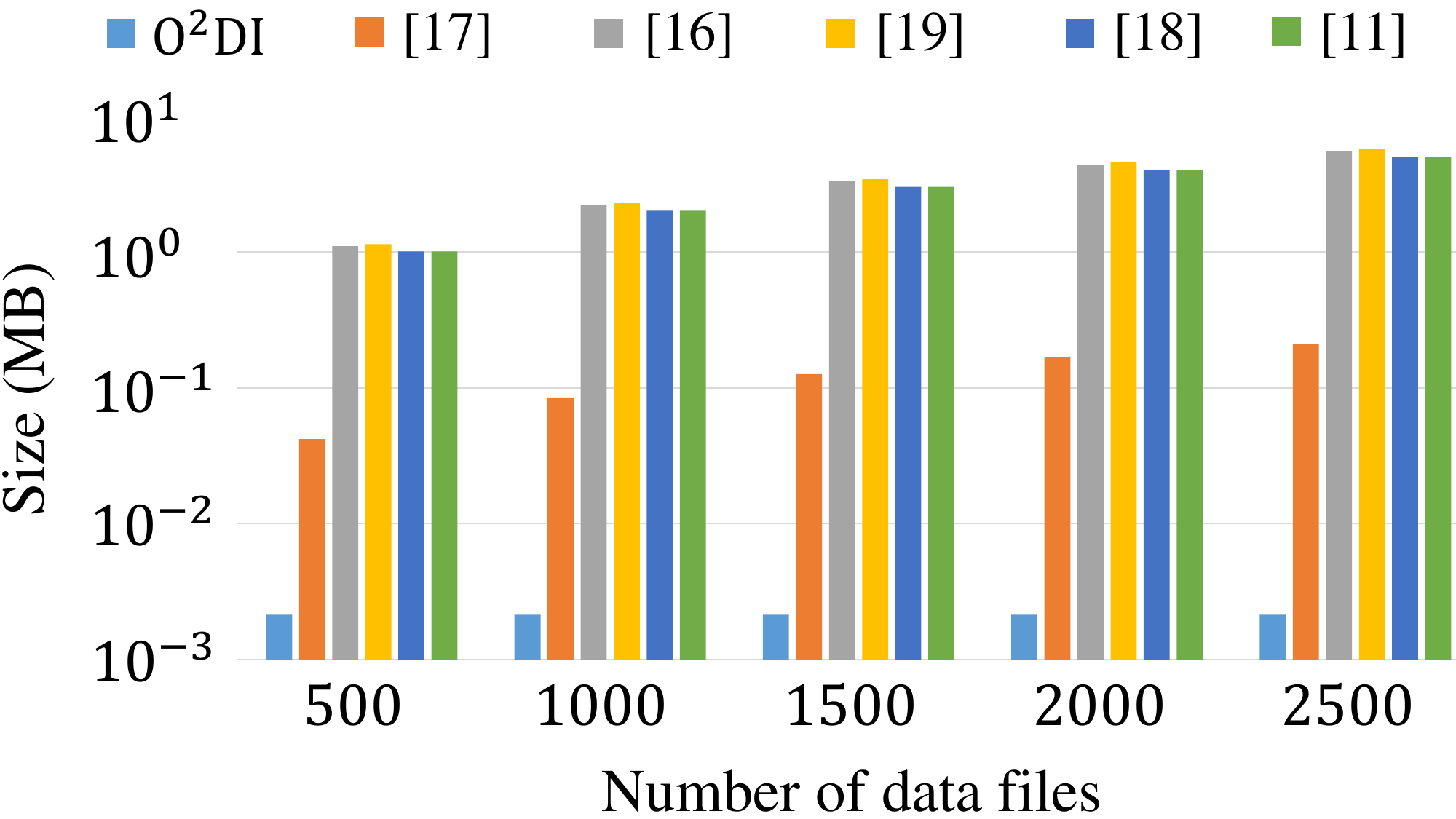} \captionsetup{labelformat=empty} &
		\includegraphics[width=1.65in]{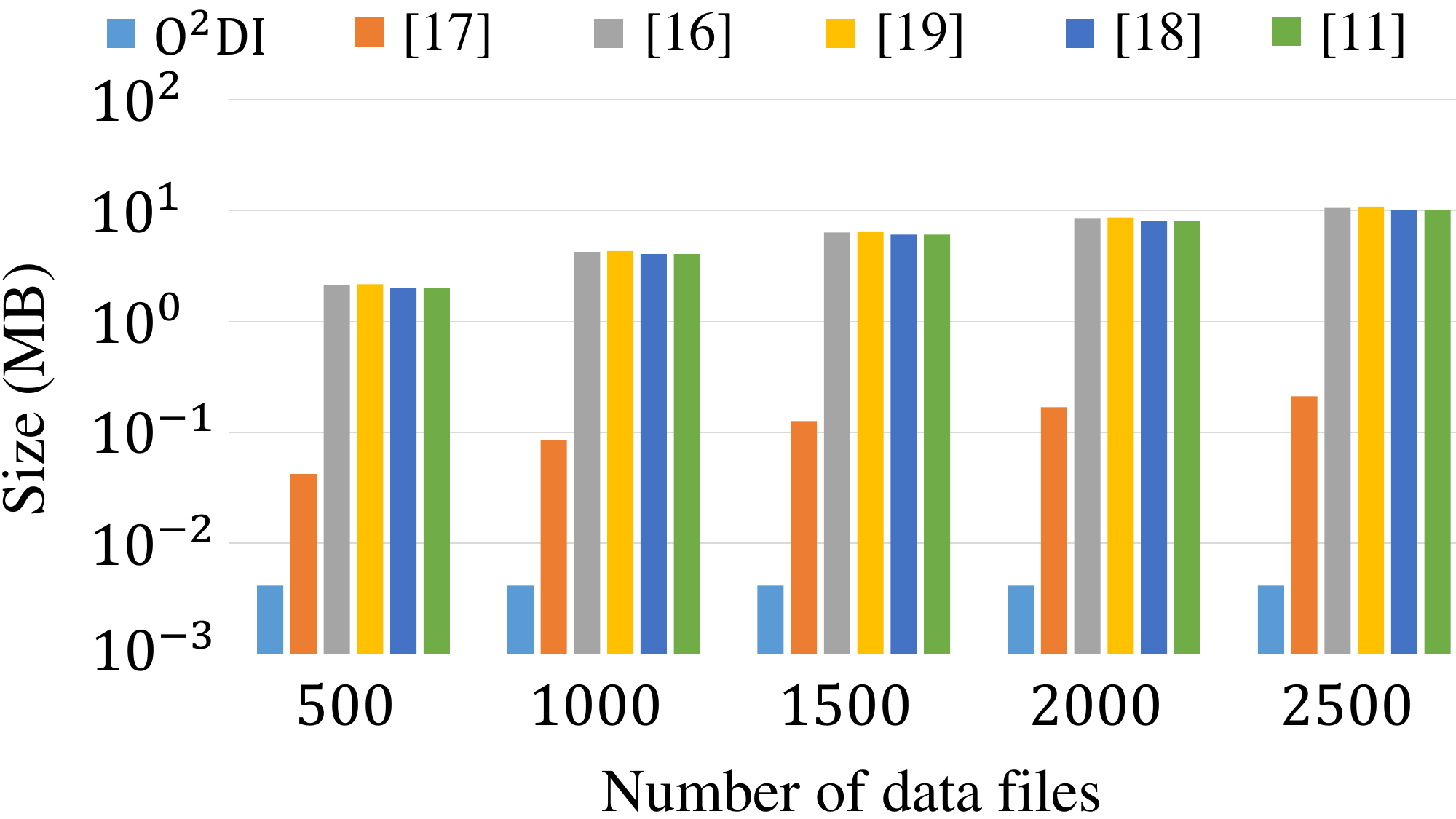} \captionsetup{labelformat=empty}&
		\includegraphics[width=1.65in]{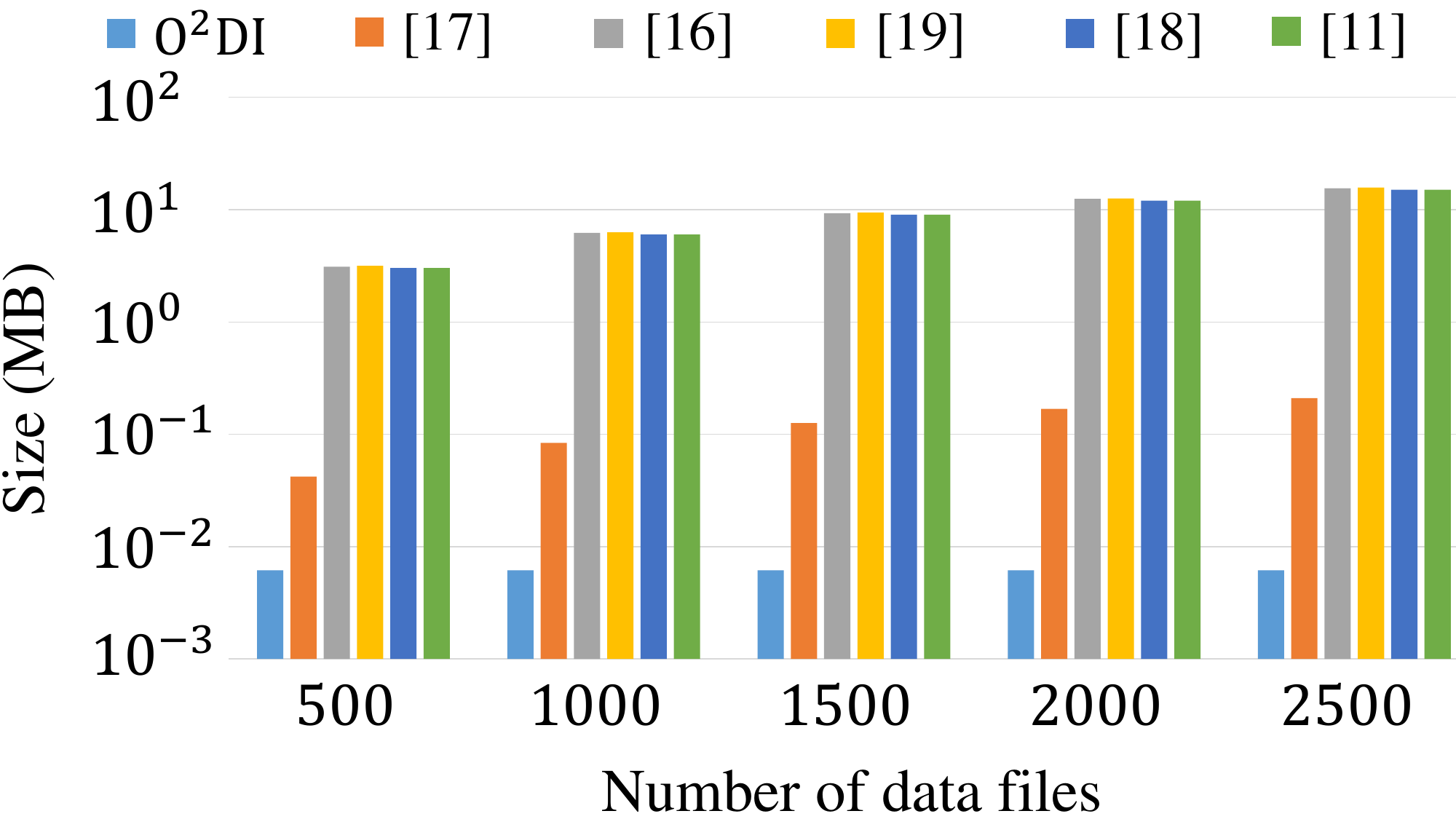} \captionsetup{labelformat=empty}&
		\includegraphics[width=1.65in]{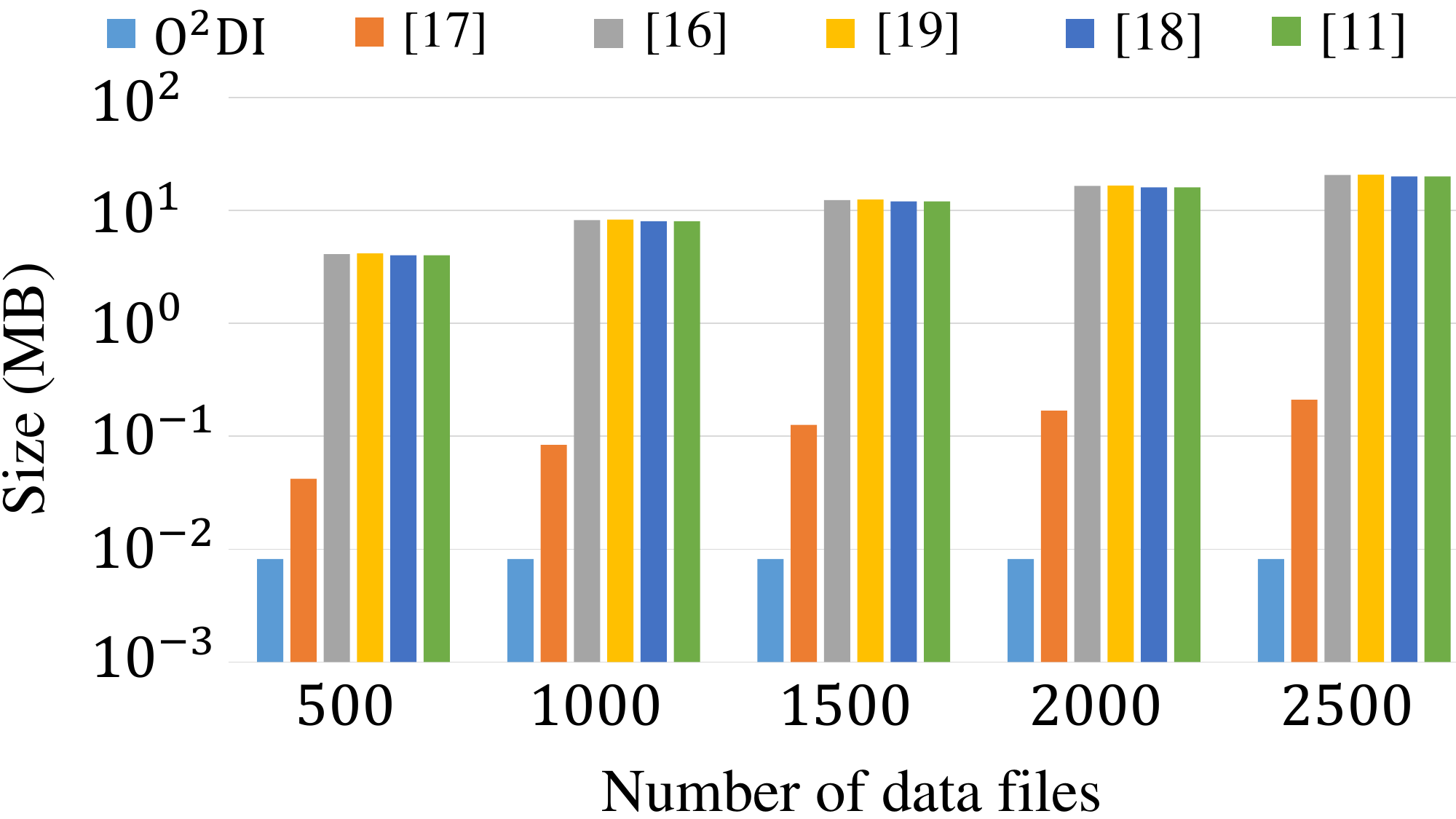} \\
		\scalebox{0.8}{(a) $I=100$} & \scalebox{0.8}{(b) $I=200$} &  \scalebox{0.8}{(c) $I=300$} & \scalebox{0.8}{(d) $I=400$}\\
		
	\end{tabular}
	\caption{The communication overhead from the  app vendor to ESs for transmitting challenges. }
	\label{ChallengelenTime}
\end{figure*}

\setcounter{figure}{4} 
\begin{figure*}[t]
	\multifig
	\centering
	\begin{tabular}{cccc}
		\figtarget & \figtarget & \figtarget& \figtarget \\
		\includegraphics[width=1.65in]{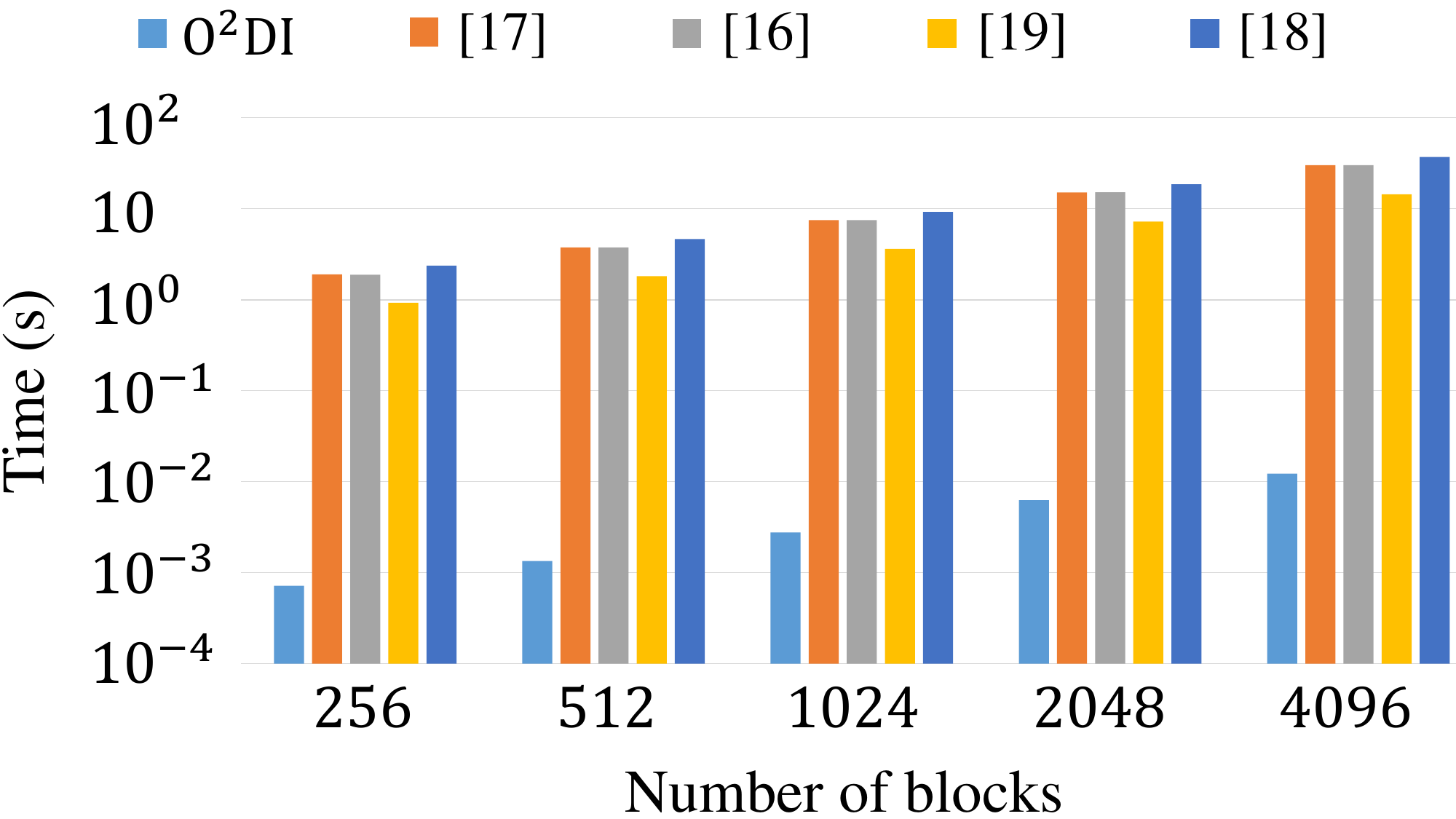} \captionsetup{labelformat=empty} &
		\includegraphics[width=1.65in]{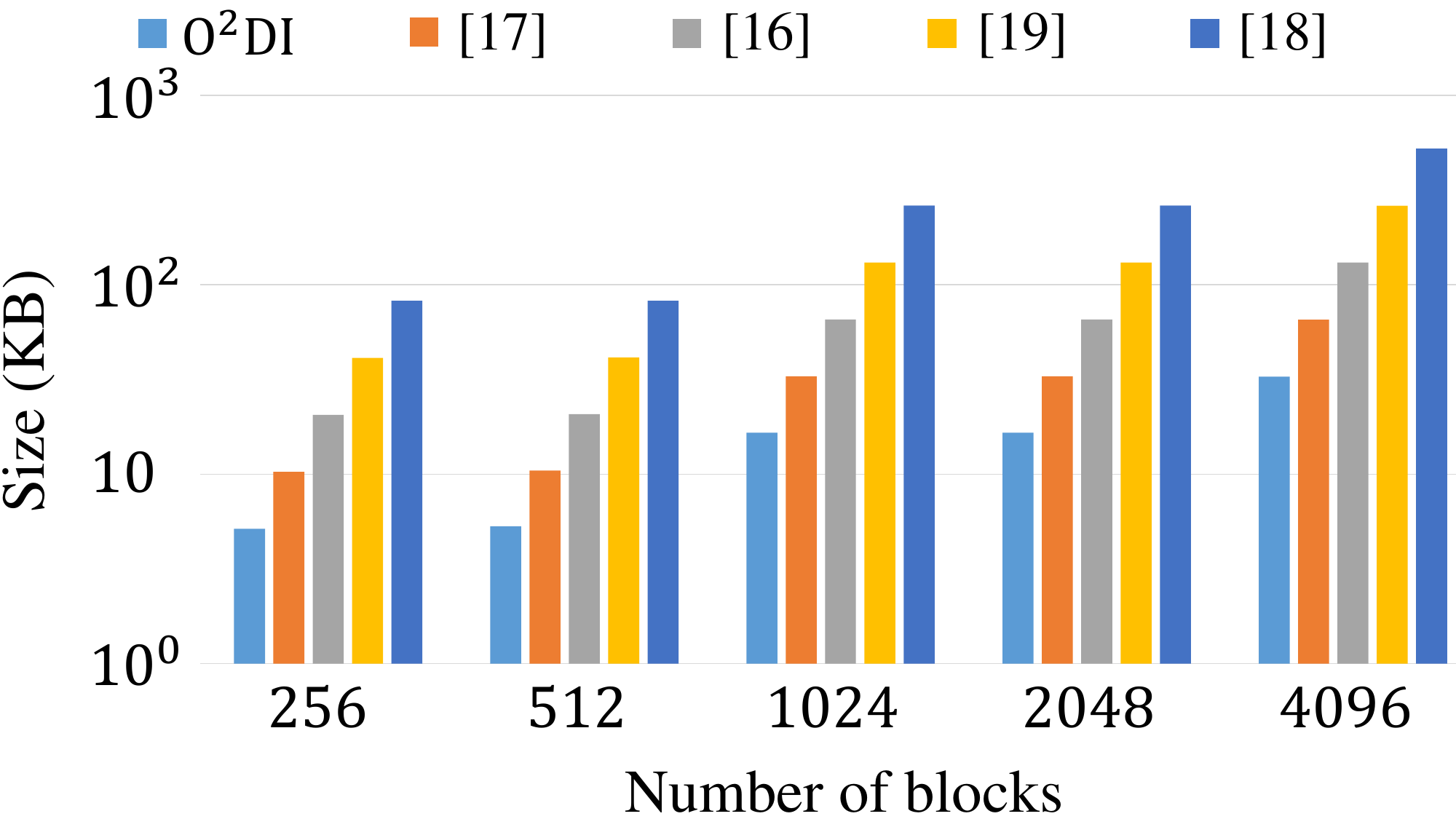} \captionsetup{labelformat=empty}&
		\includegraphics[width=1.65in]{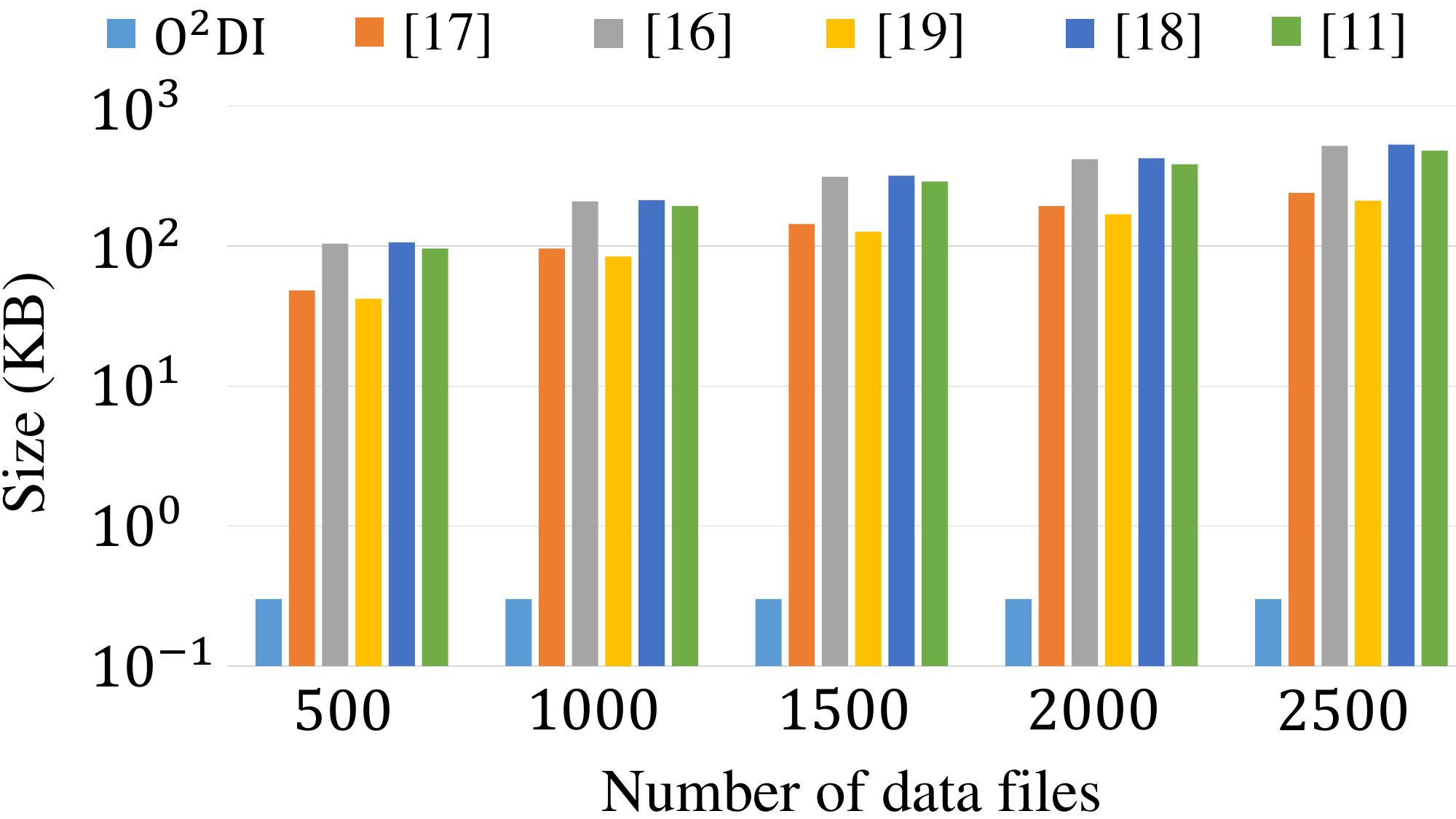} \captionsetup{labelformat=empty}&
		\includegraphics[width=1.65in]{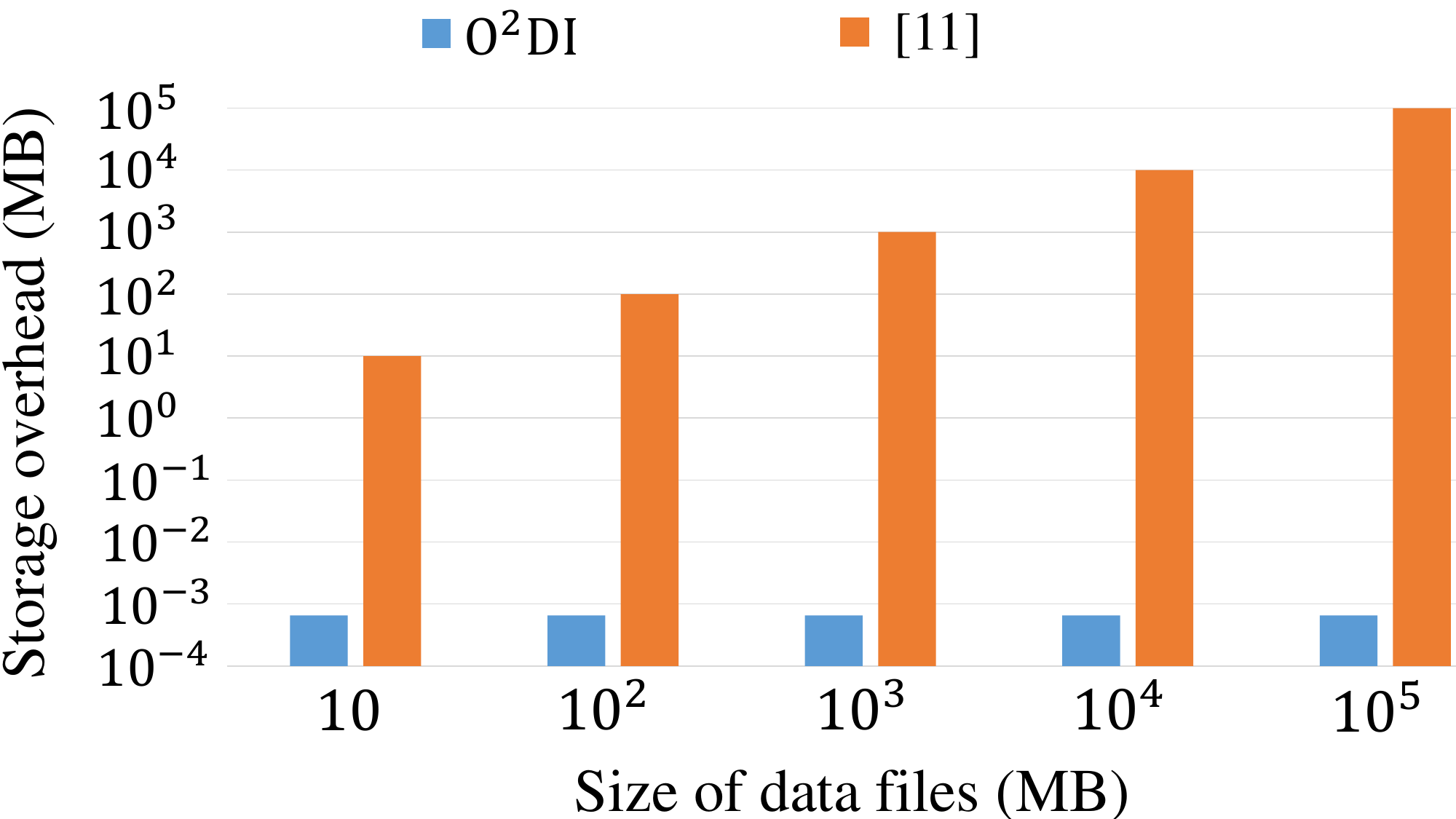} \\
		(a)  & (b) &  (c) & (d) \\
		
	\end{tabular}
	\caption{(a) tag generation time; (b) size of a tag; (c) size of integrity proofs; (d) storage overhead on the app vendor. }
	\label{ChallengeGenTime}
\end{figure*}


	\section{Performance Analysis}\label{uhgjnba}
	In this section, we  evaluate the efficiency of ${\text{O}^2\text{DI}}$ by comparing its execution time, storage cost, and communication overhead with approaches described in {\cite{LI,Yuu,dynamic,Blockchainnn,certificate-based}}. The comparison results are presented in terms of   both actual execution results and   asymptotic complexity. 
	
	Table \ref{taa} presents our asymptotic analysis. In this table, we accurately calculate different overhead  incurred in the use of the ${\text{O}^2\text{DI}}$ approach and the schemes presented in {\cite{LI,Yuu,dynamic,Blockchainnn,certificate-based}}. In the computation of execution time overhead ($\mathcal{O}_1$ to $\mathcal{O}_4$), we considered  the time consumption of pairing operation, $T_p$, exponential operation in $G_1$, $T_{e_1}$,  exponential operation in $G_2$, $T_{e_2}$, signature generation, $T_{Sign}$, signature verification, $T_{Vrfy}$,  hashing operation, $T_H$, multiplication operation in $\mathbb{Z}_q$, $T_{M_z}$,  multiplication operation in $G_1$, $T_{G_1}$, addition operation in $\mathbb{Z}_q$, $T_{A}$, random selection of an element from $\mathbb{Z}_q$, $T_{z\leftarrow \mathbb{Z}_q}$, and the computation of a pseudorandom function, $T_F$.   Also,  storage and communication costs ($\mathcal{O}_5$ to $\mathcal{O}_8$) are calculated according to the size of  elements in $\mathbb{Z}_q$, $G_1$,  $G_2$ ($l_{\mathbb{Z}_q}$, $l_{G_1}$, $l_{G_2}$), a digital signature size, $l_{Sign}$, and size of a hash string, $l_H$.
	
	We {implement}  ${\text{O}^2\text{DI}}$ and the schemes described in  \cite{LI,Yuu,dynamic,Blockchainnn,certificate-based} on  an Ubuntu 18.04 laptop with an Intel Core i5-2410M Processor 2.3 GHz, 6 GB RAM by employing the libraries python Pairing-Based Cryptography (pyPBC) 
	\footnote{https://github.com/debatem1/pypbc.}
	 and  hashlib 
	 \footnote{https://docs.python.org/3/library/hashlib.html\# module-hashlib.}.  {In this section, we assumed that the number of data files to be verified, $N$, is   ranged between   $500$ and $2500$, the number of challenged blocks considered in the verification process, $|I|$,  is ranged from  $100$ to $400$, and the number of blocks  in each data file, $\ell$, is considered between $256$ and  $4096$.}
	Furthermore, we  employ Type A pairings and the SHA-{{3}} algorithm.  Type A pairings are constructed on the curve $y^2=x^3+x$ over the finite field $\mathbb{F}_n$, where $n$ is a prime number such that $n\mathop  \equiv \limits^4 3$ \cite{562}. Remember the algorithm $\mathcal{G}$ introduced in Section \ref{Preliminaries}.  We observed that $(\lambda,q,G_1,G_2,\hat{e})\leftarrow\mathcal{G}(1^{\lambda})$, for a security parameter $\lambda$. In this case,   $G_1$ and $G_2$ are  $q$-order subgroups of  $E(\mathbb{F}_n)$ and $E(\mathbb{F}_{n^{2}})$, respectively, and $q$ is a divisor of $n+1$ \cite{562}. In this work, we assume that   $n$ and  $q$ are  $512$-bit and $160$-bit prime numbers, respectively.

	\subsection{Execution time}
	{Figure \ref{ChallengeGenTimee} shows the actual  running time in the challenge generation process. We observe that ${\text{O}^2\text{DI}}$ reduces the time consumption in this stage  significantly. In addition to the challenge generation, we see that ${\text{O}^2\text{DI}}$ lightens the computational burden on ESs in the integrity proof generation process. Indeed, as shown in Figure \ref{ProofGenTime}, we see that    ${\text{O}^2\text{DI}}$ can make this process  more than $100$ times faster than existing approaches.  Figure \ref{IntCheckTime} demonstrates the execution overhead on the app vendor for checking  the correctness of integrity proofs. From the figure, one concludes that  ${\text{O}^2\text{DI}}$  is significantly more efficient than existing approaches in this stage as well.  Moreover, Figure \ref{ChallengeGenTime} part (a) shows that in ${\text{O}^2\text{DI}}$ the tag generation process  is dramatically faster than other schemes.

		\subsection{Communication overhead and storage overhead}
		Figure \ref{ChallengelenTime}   shows  the communication overhead from the app vendor  to ESs for transmitting  a challenge. We see that as in our proposed approach the size of a challenge is very low, the communication overhead is significantly reduced in comparison to other approaches. Also, from Part (b) of Figure \ref{ChallengeGenTime}, we observe that the size of tags generated by the ${\text{O}^2\text{DI}}$ approach is significantly low  compared to  the ones constructed according to other approaches.  Part (c) of the figure demonstrates that  ${\text{O}^2\text{DI}}$ can dramatically reduce  the size of integrity proofs. Finally, from  Figure \ref{ChallengeGenTime} part (d),  we see that our  ${\text{O}^2\text{DI}}$  is considerably more efficient than digital signature-based  approaches. Indeed, as the entirety of  data files is required in the  verification process based on digital signatures, app vendors have to locally store the data files. It significantly increases  storage costs.   Fortunately, this problem has been resolved in our ${\text{O}^2\text{DI}}$ effectively.

\section{Conclusion}\label{c1c}
In this paper, we designed a novel online/offline remote data inspection approach called ${\text{O}^2\text{DI}}$ for distributed edge computing environments. The proposed approach achieves the following: (1) batch remote data verification; (2)  very efficient computational, storage, and I/O overhead; (3) efficient corrupted data localization; (4) corrupted data repair.  We also demonstrated how to deploy our  ${\text{O}^2\text{DI}}$ in      edge computing  environments. We presented the security definition and  proved that
our proposed approach is  secure in the random oracle model. Our performance analysis showed that our approach  is much more efficient than state-of-the-art approaches, and it is    practical in actual applications. 
 
\bibliographystyle{ieeetran}
\bibliography{references}

\ifCLASSOPTIONcaptionsoff
\newpage
\fi
%
\begin{IEEEbiography}[{\includegraphics[width=1in,height=1.25in,clip,keepaspectratio]{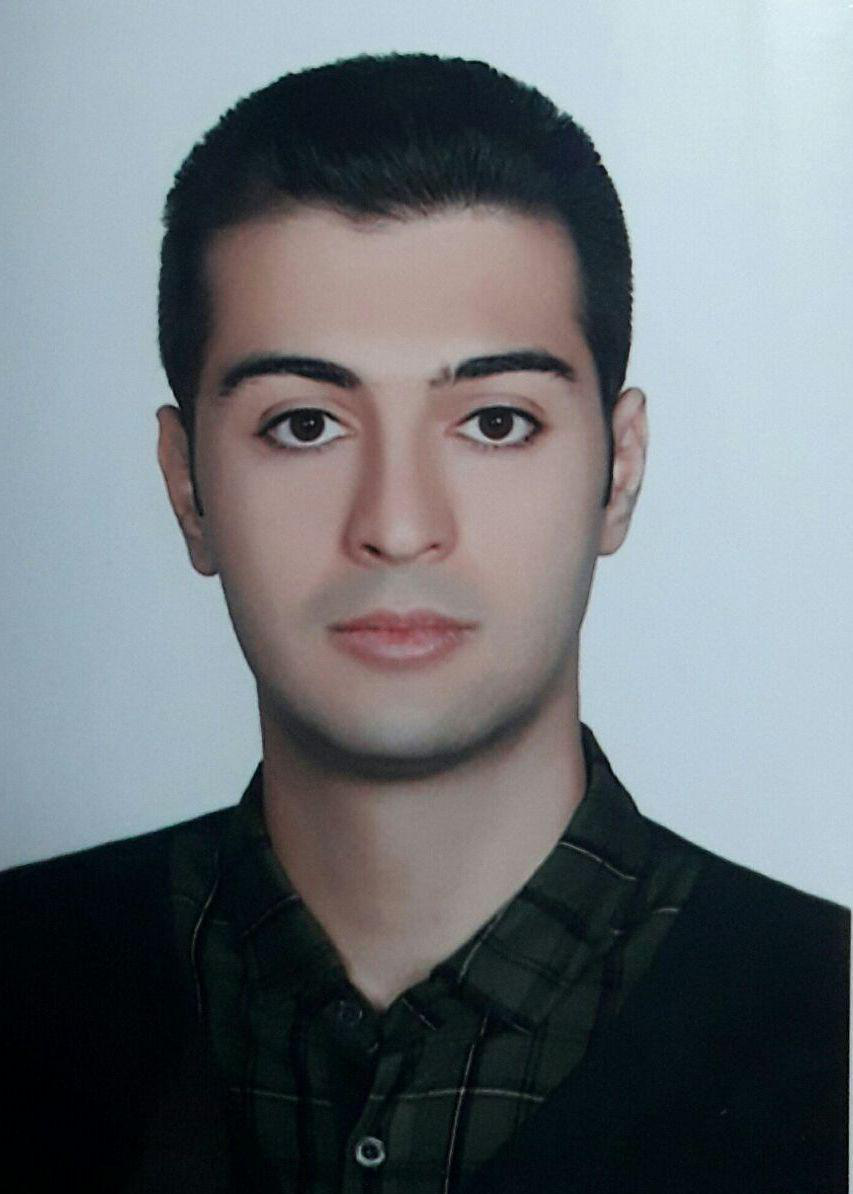}}]{Mohammad Ali}
	received the M.Sc. degree in applied mathematics, in 2016, and the Ph.D. degree in mathematics from Amirkabir University of Technology, Tehran, Iran, in 2020. He is currently an Assistant Professor with the Department
	of Mathematics and Computer Science, Amirkabir University of Technology. His research interests include provable security cryptography, post quantum cryptography,  cloud Security, and IoT  security.
	
\end{IEEEbiography}

\begin{IEEEbiography}[{\includegraphics[width=1in,height=1.25in,clip,keepaspectratio]{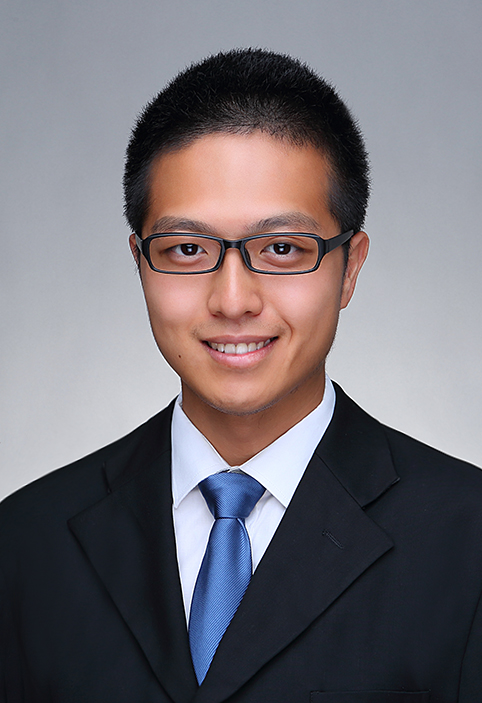}}]{Ximeng Liu}
	received the B.Sc. degree in electronic engineering from Xidian University, Xi’an,
	China, in 2010, and the Ph.D. degree in cryptography from Xidian University, China, in 2015.
	He is currently a Full Professor with the College
	of Mathematics and Computer Science, Fuzhou
	University. He was also a Research Fellow with the
	School of Information System, Singapore Management University, Singapore. He has published
	more than 100 papers on the topics of cloud security and big data security, including papers in the IEEE Transaction on Computers, the IEEE Transaction on Industrial Informatics, the IEEE
	Transaction on Dependable Secure Computing, the IEEE Transaction on Service Computing, and the IEEE Internet of Things Journal. His
	research interests include cloud security, applied cryptography, and big data
	security. He is a member of ACM and CCF. He was awarded the Minjiang
	Scholars Distinguished Professor, Qishan Scholars in Fuzhou University, and
	ACM SIGSAC China Rising Star Award, in 2018. He served as a program
	committee for several conferences, such as the 17th IEEE International	Conference on Trust, Security and Privacy in Computing and Communications, 2017 IEEE Global Communications Conference, and 2016 IEEE Global Communications Conference. He served as the Lead Guest Editor 	for Wireless Communications and Mobile Computing.
\end{IEEEbiography}

\end{document}